\documentclass{IEEEtran}
\usepackage[latin9]{inputenc}
\usepackage{amsmath}
\usepackage{amsthm}
\usepackage{graphicx}

\makeatletter
\theoremstyle{plain}
\newtheorem{thm}{\protect\theoremname}
\theoremstyle{definition}
\newtheorem{defn}[thm]{\protect\definitionname}
\theoremstyle{plain}
\newtheorem{cor}[thm]{\protect\corollaryname}
\theoremstyle{plain}
\newtheorem{lem}[thm]{\protect\lemmaname}
\theoremstyle{remark}
\newtheorem{claim}[thm]{\protect\claimname}

\usepackage{amsfonts}
\usepackage{cite}

\makeatother

\providecommand{\claimname}{Claim}
\providecommand{\corollaryname}{Corollary}
\providecommand{\definitionname}{Definition}
\providecommand{\lemmaname}{Lemma}
\providecommand{\theoremname}{Theorem}

\begin{document}
\global\long\def\var{\mathbb{V}}
 \global\long\def\bias{\mathbb{B}}
 \global\long\def\ft#1{\tilde{\mathbf{f}}_{#1,h_{#1}}}
 \global\long\def\gt{\tilde{\mathbf{G}}_{h_{1},h_{2}}}
 \global\long\def\gtm{\tilde{\mathbf{G}}_{h_{X},h_{Y}}}
 \global\long\def\gh{\tilde{\mathbf{G}}_{h}}

\global\long\def\g#1{\tilde{\mathbf{G}}_{#1}}
 \global\long\def\gtl{\tilde{\mathbf{G}}_{h_{X}(l),h_{Y}(l)}}

\global\long\def\S{\mathcal{S}}

\global\long\def\ez{\mathbb{E}_{\mathbf{Z}}}
 \global\long\def\bE{\mathbb{E}}
 \global\long\def\et#1{\tilde{\mathbf{e}}_{#1,h_{#1}}}
 \global\long\def\fhatl#1#2{\tilde{\mathbf{f}}_{#1,h(#2)}}
 \global\long\def\ftl#1{\tilde{\mathbf{f}}_{#1,h(l)}}

\global\long\def\ett#1#2{\tilde{\mathbf{e}}_{#1#2,h_{#1},h_{#2}}}

\global\long\def\dist#1{\mathbf{\rho}_{#1,k_{#1}}}
 \global\long\def\fh#1{\hat{\mathbf{f}}_{#1,k_{#1}}}
 \global\long\def\ghat{\hat{\mathbf{G}}_{k_{1},k_{2}}}
 \global\long\def\gk{\hat{\mathbf{G}}_{k}}
 \global\long\def\ehat#1{\hat{\mathbf{e}}_{#1,k_{#1}}}

\global\long\def\fb#1{\bar{\mathbf{f}}_{#1,k_{#1}}}
 \global\long\def\fbp#1{\bar{\mathbf{f}}_{#1,k_{#1}+1}}
 \global\long\def\fhl#1#2{\hat{\mathbf{f}}_{#1,k(#2)}}
 \global\long\def\ehatl#1#2{\bar{\mathbf{e}}_{#1,k(#2)}}
 \global\long\def\gtay#1#2#3{\mathbf{#1}_{#2}^{(#3)}}

\onecolumn

\title{Ensemble Estimation of Distributional Functionals via $k$-Nearest
Neighbors}

\author{Kevin R. Moon$^{1}$, Kumar Sricharan$^{2}$, Alfred O. Hero III$^{3}$
\thanks{This work was partially supported by ARO grant W911NF-15-1-0479.}\\
$^{1}$Genetics Department and Applied Math Program, Yale University\\
$^{2}$Xerox PARC\\
$^{3}$EECS Department, University of Michigan\textsuperscript{}}
\maketitle
\begin{abstract}
The problem of accurate nonparametric estimation of distributional
functionals (integral functionals of one or more probability distributions)
has received recent interest due to their wide applicability in signal
processing, information theory, machine learning, and statistics.
In particular, $k$-nearest neighbor (nn) based methods have received
a lot of attention due to their adaptive nature and their relatively
low computational complexity. We derive the mean squared error (MSE)
convergence rates of leave-one-out $k$-nn plug-in density estimators
of a large class of distributional functionals without boundary correction.
We then apply the theory of optimally weighted ensemble estimation
to obtain weighted ensemble estimators that achieve the parametric
MSE rate under assumptions that are competitive with the state of
the art. The asymptotic distributions of these estimators, which are
unknown for all other $k$-nn based distributional functional estimators,
are also presented which enables us to perform hypothesis testing. 
\end{abstract}

\section{Introduction}

Information measures such as entropy, information divergence, and
mutual information, are useful in many applications in signal processing,
information theory, machine learning, and statistics. These information
measures belong to a larger class of functionals known as distributional
functionals, defined as integral functionals of one or more probability
distributions. Distributional functionals have been used in applications
such as Bayes error rate estimation \cite{berisha2014bound,gliske2015intrinsic,hashlamoun1994bound,avi1996bound,chernoff1952measure,moon2015Bayes},
the two sample test \cite{henze1999multivariate}, estimating the
decay rates of error probabilities \cite{cover2012elements}, clustering
\cite{banerjee2005clustering,dhillon2003cluster,lewi2006real}, intrinsic
dimension estimation\cite{carter2010local,moon2015partI}, feature
selection and classification \cite{bruzzone1995feature,guorong1996feature,sakate2014variable},
image segmentation \cite{liu2014segment}, extending machine learning
algorithms to distributional features \cite{poczos2011estimation,szabo2014distribution,oliva2013distribution,moon2015partII},
steganography \cite{korzhik2015steganographic}, and structure learning
\cite{chow1968approximating,moon2017information}. 

We consider the problem of nonparametric estimation of these distributional
functionals from a finite population of i.i.d. samples drawn from
each $d$-dimensional distribution without any knowledge of the boundary
of the densities' support set. We derive the mean squared error (MSE)
convergence rates of leave-one-out $k$-nearest neighbor (nn) plug-in
density estimators. We then apply the general theory of optimally
weighted ensemble estimation developed in \cite{moon2016arxiv,moon2016isit,sricharan2013ensemble}
to obtain weighted ensemble estimators that achieve the parametric
MSE convergence rate of $O\left(1/N\right)$ when the densities are
sufficiently smooth, where $N$ is the sample size. We also derive
the asymptotic distribution of the weighted ensemble estimators.

For brevity, we focus on estimating functionals of two distributions
(referred to as divergence functionals) in this paper. However, our
methods can be easily extended to functionals of any finite number
of distributions.

Several previous works have explored $k$-nn estimators for distributional
functionals. Poczos and Schneider \cite{poczos2011estimation} proved
that a fixed $k$-nn estimator with bias correction is weakly consistent
for Renyi-$\alpha$ and other similar divergences. Wang et al \cite{wang2009divergence}
provided a $k$-nn based estimator for the Kullback-Leibler divergence
while Gao et al \cite{gao2017density} proved the consistency of local
likelihood density estimators with $k$-nn bandwidths for polynomials
of a single distribution. However, none of these works study the MSE
convergence rates nor the asymptotic distribution of their estimators.

More recent work has focused on the convergence rates of $k$-nn based
estimators of distributional functionals. Gao et al \cite{gao2017demystifying}
showed that popular $k$-nn based Shannon entropy \cite{kozachenko1987sample}
and Shannon mutual information \cite{kraskov2004estimating} estimators
achieve the parametric MSE rate when the dimension of each of the
random variables is less than 3. Singh and Poczos \cite{singh2016finite}
derive the convergence rates for fixed $k$-nn estimators of specific
distributional functionals where a bias correction term is known and
when the densities' support set contains no boundaries. 

Ensemble techniques \cite{moon2016arxiv,moon2016isit,sricharan2013ensemble}
have previously been applied to $k$-nn based estimators of some distributional
functionals to obtain estimators that achieve the parametric rate
when the densities are sufficiently smooth. Noshad et al \cite{noshad2017direct}
and Wisler et al \cite{wisler2017direct} applied ensemble techniques
to $k$-nn based direct estimators of $f$-divergence functionals.
Moon and Hero \cite{moon2014isit} applied ensemble techniques to
$k$-nn plug-in estimators of $f$-divergences and applied ensemble
techniques to obtain an estimator that achieves the parametric rate
when the densities' support set is compact and contains no boundaries,
or when boundary correction is applied. However, our assumptions on
the smoothness of the densities are less strict than required for
some of these estimators \cite{moon2014isit,noshad2017direct} and
we consider different boundary conditions on the densities' support
set. Additionally, our techniques can be applied to a larger class
of distributional functionals which includes the $L^{2}$ divergence
and general entropies whereas the work in \cite{noshad2017direct,moon2014isit,wisler2017direct}
is limited to $f$-divergence functionals (functionals of the likelihood
ratio between two densities). Furthermore, while Moon and Hero \cite{moon2014nips}
derive the asymptotic distribution for the plug-in estimators defined
in \cite{moon2014isit}, the asymptotic distributions of the estimators
in \cite{wisler2017direct,noshad2017direct} are unknown. In contrast,
we obtain the asymptotic distribution of our estimators under much
less strict assumptions on the densities and the functional compared
to the work in \cite{moon2014nips}.

Many other approaches for distributional functional estimation have
also been examined including methods based on kernel density estimators
(KDE) \cite{moon2016arxiv,moon2016isit,krishnamurthy2014divergence,kandasamy2015nonparametric,singh2014exponential,singh2014renyi,sricharan2013ensemble}
and convex risk minimization \cite{nguyen2010div}. While all of these
works define estimators that can achieve the parametric MSE rate,
these methods are generally more computationally intensive than $k$-nn
based methods and some of them require explicit knowledge of the densities'
support set boundary \cite{krishnamurthy2014divergence,kandasamy2015nonparametric,singh2014exponential,singh2014renyi,sricharan2013ensemble}.

Finally, Gao et al \cite{gao2015efficient} showed that $k$-nn or
KDE based approaches underestimate the mutual information when the
mutual information is large. As mutual information increases, the
dependencies between random variables becomes more deterministic which
results in less smooth densities. This is consistent with the work
in \cite{kandasamy2015nonparametric,krishnamurthy2014divergence,moon2016arxiv,moon2016isit,moon2014isit,singh2014renyi,singh2014exponential}
and this work which require the densities to be smooth to achieve
the parametric rate.

The remainder of the paper is as follows. Section~\ref{sec:knn_est}
presents the divergence functional $k$-nn plug-in estimators and
the corresponding MSE convergence rates. We then apply ensemble estimation
theory to these estimators in Section~\ref{sec:ensemble} to obtain
estimators that achieve the parametric MSE rate when the densities
are sufficiently smooth. A central limit theorem is given. We then
numerically validate the estimators in Section~\ref{sec:experiments}.
All proofs are reserved for the appendices. Bold face type is used
for random variables and random vectors. The conditional expectation
given a random variable $\mathbf{Z}$ is denoted $\ez$. The variance
of a random variable is denoted $\var$ and the bias of an estimator
is denoted $\bias$.

\section{The Divergence Functional Plug-in Estimator}

\label{sec:knn_est} We focus on estimating functionals of two distributions
of the form 
\begin{equation}
G\left(f_{1},f_{2}\right)=\int g\left(f_{1}(x),f_{2}(x)\right)f_{2}(x)dx,\label{eq:div_functional}
\end{equation}
where $f_{1}$ and $f_{2}$ are smooth $d$-dimensional probability
densities and $g(t_{1},t_{2})$ is a smooth functional.

\subsection{The $k$-nn Plug-in Estimator}

We use a $k$-nn density plug-in estimator of the divergence functional
in (\ref{eq:div_functional}). Assume that $N_{1}$ i.i.d. samples
$\left\{ \mathbf{Y}_{1},\dots,\mathbf{Y}_{N_{1}}\right\} $ are available
from $f_{1}$ and $N_{2}$ i.i.d. samples $\left\{ \mathbf{X}_{1},\dots,\mathbf{X}_{N_{2}}\right\} $
are available from $f_{2}$. Let $M_{1}=N_{1}$, $M_{2}=N_{2}-1$,
and $k_{i}\leq M_{i}$. Denote the distance of the $k_{1}$th nearest
neighbor of the sample $\mathbf{Y}_{i}$ in $\left\{ \mathbf{X}_{1},\dots,\mathbf{X}_{N_{2}}\right\} $
as $\dist 1(i)$. Similarly, denote the distance of the $k_{2}$th
nearest neighbor of the sample $\mathbf{X}_{i}$ in $\left\{ \mathbf{X}_{1},\dots,\mathbf{X}_{N_{2}}\right\} \backslash\left\{ \mathbf{X}_{i}\right\} $
as $\dist 2(i)$. The standard $k$-nn density estimator is~\cite{loftsgaarden1965knn}
\[
\fh i(\mathbf{X}_{j})=\frac{k_{i}}{M_{i}c_{d}\dist i^{d}(j)},
\]
where $c_{d}$ is the volume of a $d$-dimensional unit ball. The
functional $G(f_{1},f_{2})$ is estimated as 
\[
\ghat=\frac{1}{N_{2}}\sum_{i=1}^{N_{2}}g\left(\fh 1(\mathbf{X}_{i}),\fh 2(\mathbf{X}_{i})\right).
\]

\subsection{Convergence Rates}

We derive the MSE convergence rate of our estimators in terms of the
H\"{o}lder condition:
\begin{defn}
[H\"{o}lder Class] \label{def:holder}Let $\mathcal{X}\subset\mathbb{R}^{d}$
be a compact space. For $r=(r_{1},\dots,r_{d})$, $r_{i}\in\mathbb{N}$,
define $|r|=\sum_{i=1}^{d}r_{i}$ and $D^{r}=\frac{\partial^{|r|}}{\partial x_{1}^{r_{1}}\dots\partial x_{d}^{r_{d}}}$.
The H\"{o}lder class $\Sigma(s,K)$ of functions on $L_{2}(\mathcal{X})$
consists of the functions $f$ that satisfy 
\[
\left|D^{r}f(x)-D^{r}f(y)\right|\leq K\left\Vert x-y\right\Vert ^{s-\lfloor s\rfloor},
\]
 for all $x,\,y\in\mathcal{X}$ and for all $r$ s.t. $|r|\leq\left\lfloor s\right\rfloor $. 
\end{defn}
Consider the following assumptions:
\begin{itemize}
\item $(\mathcal{B}.1)$: Assume there exist constants $\epsilon_{0},\epsilon_{\infty}$
such that $0<\epsilon_{0}\leq f_{i}(x)\leq\epsilon_{\infty}<\infty,\,\forall x\in\mathcal{S}.$ 
\item $(\mathcal{B}.2)$: Assume that the densities $f_{i}\in\Sigma(s,K)$
in the interior of $\mathcal{S}$ with $s\geq2$ and $r=\lfloor s\rfloor$. 
\item $(\mathcal{B}.3)$: Assume that $g$ has an infinite number of mixed
derivatives. 
\item $(\mathcal{B}.4$): Assume that $\left|\frac{\partial^{k+l}g(x,y)}{\partial x^{k}\partial y^{l}}\right|$,
$k,l=0,1,\ldots$ are strictly upper bounded for $\epsilon_{0}\leq x,y\leq\epsilon_{\infty}$. 
\item $(\mathcal{B}.5)$: Assume that the densities' support set is $\mathcal{S}=[0,1]^{d}$. 
\end{itemize}
These assumptions enable us to obtain the bias results for the $k$-nn
plug in estimator $\ghat$. Assumption $\mathcal{B}.3$ is used to
obtain the bias convergence rates without knowledge of the boundary
of the densities' support set. This assumption is not overly restrictive
as most divergence functionals of interest are infinitely differentiable.
Those functionals that are not infinitely differentiable are typically
not differentiable everywhere (e.g. the total variation distance and
the Bayes error) which violates the assumptions of current nonparametric
estimators that achieve the parametric rate. Assumption $\mathcal{B}.5$
is used to handle the boundary bias of the $k$-nn estimators. In
particular, the proof derives the bias contribution of points that
are near the flat ``walls'' of the cube and near the corners. Thus
our results still hold for rotated and stretched or compressed support
sets. It is also likely that our results can be extended to other
support sets with relatively smooth boundaries and some sharp corners.
In contrast, the theory developed in \cite{noshad2017direct,moon2014isit,wisler2017direct}
applies when the densities' support set contains no boundaries (e.g.
the surface of the hypersphere) \cite{moon2014isit}, the densities
decay to zero near the support set boundary \cite{wisler2017direct},
or the derivatives of the densities decay to zero near the support
set boundary \cite{noshad2017direct}. 

The following theorem on the bias of the plug-in estimator follows
under assumptions $\mathcal{B}.1-\mathcal{B}.5$. For simplicity,
assume that $N_{1}=N_{2}=N$ and $k_{1}=k_{2}=k$.
\begin{thm}
\label{thm:biasKNN} For general $g$, the bias of the plug-in estimator
$\ghat$ is of the form 
\begin{eqnarray}
\bias\left[\ghat\right] & = & \sum_{j=1}^{r}\left(\left(c_{17,1,j}+\frac{c_{17,1,j,0}}{\sqrt{k_{1}}}\right)\left(\frac{k_{1}}{N_{1}}\right)^{\frac{j}{d}}+\left(c_{17,2,j}+\frac{c_{17,2,j,0}}{\sqrt{k_{2}}}\right)\left(\frac{k_{2}}{N_{2}}\right)^{\frac{j}{d}}\right)\nonumber \\
 &  & +\sum_{j=0}^{r}\sum_{\substack{i=0\\
i+j\neq0
}
}^{r}c_{18,i,j}\left(\frac{k_{1}}{N_{1}}\right)^{\frac{i}{d}}\left(\frac{k_{2}}{N_{2}}\right)^{\frac{j}{d}}\nonumber \\
 &  & +O\left(\frac{1}{\sqrt{k_{1}k_{2}}}+\frac{1}{k_{1}}+\frac{1}{k_{2}}+\max\left(\frac{k_{1}}{N_{1}},\frac{k_{2}}{N_{2}}\right)^{\frac{\min(s,d)}{d}}\right).\label{eq:bias1KNN}
\end{eqnarray}
Furthermore, if $g(x,y)$ has $m,\,l$-th order mixed derivatives
$\frac{\partial^{m+l}g(x,y)}{\partial x^{m}\partial y^{l}}$ that
depend on $x,y$ only through $x^{\alpha}y^{\beta}$ for some $\alpha,\beta\in\mathbb{R}$,
then for any positive integer $\nu\geq0$, the bias is of the form
\begin{eqnarray}
\bias\left[\ghat\right] & = & \sum_{j=0}^{\lfloor s\rfloor}\sum_{\substack{i=0\\
i+j\neq0
}
}^{\lfloor s\rfloor}c_{18,i,j}\left(\frac{k_{1}}{N_{1}}\right)^{\frac{i}{d}}\left(\frac{k_{2}}{N_{2}}\right)^{\frac{j}{d}}+O\left(\max\left(\frac{k_{1}}{N_{1}},\frac{k_{2}}{N_{2}}\right)^{\frac{\min(s,d)}{d}}+\frac{1}{\min(k_{1},k_{2})^{\frac{2+\nu}{2}}}\right)\nonumber \\
 &  & +\sum_{m=0}^{\nu}\sum_{\substack{j=0\\
j+m\neq0
}
}^{r}\left(\frac{c_{20,1,j,m}}{k_{1}^{\frac{1+m}{2}}}\left(\frac{k_{1}}{N_{1}}\right)^{\frac{j}{d}}+\frac{c_{20,2,j,m}}{k_{2}^{\frac{1+m}{2}}}\left(\frac{k_{2}}{N_{2}}\right)^{\frac{j}{d}}\right)\nonumber \\
 &  & +\sum_{j=1}^{r}\left(c_{17,1,j}\left(\frac{k_{1}}{N_{1}}\right)^{\frac{j}{d}}+c_{17,2,j}\left(\frac{k_{2}}{N_{2}}\right)^{\frac{j}{d}}\right)\nonumber \\
 &  & +\sum_{m=0}^{\nu}\sum_{\substack{j=0\\
m+j\neq0
}
}^{\lfloor s\rfloor}\sum_{n=0}^{\nu}\sum_{\substack{i=0\\
n+i\neq0
}
}^{\lfloor s\rfloor}\frac{c_{18,i,j,m,n}}{k_{1}^{\frac{1+m}{2}}k_{2}^{\frac{1+n}{2}}}\left(\frac{k_{1}}{N_{1}}\right)^{\frac{i}{d}}\left(\frac{k_{2}}{N_{2}}\right)^{\frac{j}{d}}.\label{eq:bias2KNN}
\end{eqnarray}
\end{thm}
The following variance result requires much less strict assumptions:
\begin{thm}
\label{thm:varianceKNN}If the functional $g$ is Lipschitz continuous
in both of its arguments with Lipschitz constant $C_{g}$, then the
variance of $\ghat$ is 
\begin{equation}
\var\left[\ghat\right]=O\left(\frac{1}{N_{2}}+\frac{N_{1}}{N_{2}^{2}}\right).\label{eq:varknn}
\end{equation}
\end{thm}
From Theorems~\ref{thm:biasKNN} and \ref{thm:varianceKNN}, it is
clear that we require $k_{i}\rightarrow\infty$ and $k_{i}/N_{i}\rightarrow0$
for $\ghat$ to be unbiased. For the variance to decrease to zero,
we require $N_{2}\rightarrow\infty$ and $N_{1}/N_{2}^{2}\rightarrow0$.
The additional terms in (\ref{eq:bias2KNN}) enable us to achieve
the parametric MSE convergence rate when $s>d/2$ (similar to the
estimators in \cite{wisler2017direct}) for an appropriate choice
of $k$ values whereas the terms in (\ref{eq:bias1KNN}) require $s\geq d$
to achieve the same rate (similar to the estimators in \cite{moon2014isit,noshad2017direct}).
Moreover, the additional terms in (\ref{eq:bias2KNN}) enable us to
achieve the parametric rate for smaller values of $k$ which is more
computationally efficient. 

The Lipschitz condition on $g$ is comparable to other nonparametric
estimators of distributional functionals \cite{kandasamy2015nonparametric,krishnamurthy2014divergence,moon2016arxiv,singh2014exponential,singh2014renyi}.
Specifically, assumption $\mathcal{B}$.1 ensures that functionals
such as those for Shannon and Renyi divergences are Lipschitz on the
space $\epsilon_{0}$ to $\epsilon_{\infty}$.

From Theorem~\ref{thm:biasKNN}, the dominating terms in the bias
are $\Theta\left(\left(\frac{k_{i}}{N_{i}}\right)^{\frac{1}{d}}\right)$
and $\Theta\left(\frac{1}{k_{i}}\right)$. If no bias correction is
made, the optimal choice of $k_{i}$ that minimizes the MSE is 
\[
k_{i}^{*}=\Theta\left(N_{i}^{\frac{1}{d+1}}\right).
\]
This results in a dominant bias term of order $\Theta\left(N_{i}^{\frac{-1}{d+1}}\right)$,
which is large whenever $d$ is not small.

\subsection{Proof Sketches of Theorems~\ref{thm:biasKNN} and \ref{thm:varianceKNN}}

The proof of the bias result uses a conditioning argument on the $k$-nn
distances by viewing the $k$-nn estimator as a kernel density estimator
with uniform kernel and random bandwidth. This allows us to leverage
some KDE plug-in estimator proof techniques. For fixed bandwidth (i.e.
$k$-nn distance), we then consider separately the cases where the
$k$-nn ball is contained within the support and when it intersects
the boundary of the support. See Appendix~\ref{sec:biasProofKNN}
for the full proof.

The proof of the variance result uses the Efron-Stein inequality,
which becomes complicated due to the dependencies between different
$k$-nn neighborhoods. Thus we analyze the possible effects on the
$k$-nn graph when one sample is allowed to differ in order to use
the Efron-Stein inequality. See Appendix~\ref{sec:varProofKNN} for
the full proof of Theorem~\ref{thm:varianceKNN}.

\section{Weighted Ensemble Estimation}

\label{sec:ensemble}The $k$-nn plug-in estimator $\ghat$ in Section~\ref{sec:knn_est}
has slowly decreasing bias when the dimension of the data is not small.
By applying the theory of optimally weighted ensemble estimation derived
in \cite{moon2016isit,moon2016arxiv}, we can take a weighted sum
of an ensemble of estimators where the weights are chosen to reduce
the bias.

We simplify the bias expressions in Theorem~\ref{thm:biasKNN} by
assuming that $N_{1}=N_{2}=N$ and $k_{1}=k_{2}=k$. Define $\gk:=\hat{\mathbf{G}}_{k,k}$.
\begin{cor}
\label{cor:biasKNN} For general $g$, the bias of the plug-in estimator
$\gk$ is given by 
\begin{eqnarray*}
\bias\left[\gk\right] & = & \sum_{j=1}^{r}\left(c_{21,1,j}+\frac{c_{21,2,j}}{\sqrt{k}}\right)\left(\frac{k}{N}\right)^{\frac{j}{d}}+O\left(\frac{1}{k}+\left(\frac{k}{N}\right)^{\frac{\min(s,d)}{d}}\right).
\end{eqnarray*}
If $g(x,y)$ has $m,\,l$-th order mixed derivatives $\frac{\partial^{m+l}g(x,y)}{\partial x^{m}\partial y^{l}}$
that depend on $x,y$ only through $x^{\alpha}y^{\beta}$ for some
$\alpha,\beta\in\mathbb{R}$, then for any positive integer $\nu\geq2$,
the bias is of the form
\[
\bias\left[\gk\right]=\sum_{j=1}^{r}c_{22,j}\left(\frac{k}{N}\right)^{\frac{j}{d}}+\sum_{m=0}^{\nu}\sum_{\substack{j=0\\
j+m\neq0
}
}^{r}\frac{c_{22,j,m}}{k^{\frac{1+m}{2}}}\left(\frac{k}{N}\right)^{\frac{j}{d}}+O\left(\frac{1}{k^{\frac{\nu}{2}}}+\left(\frac{k}{N}\right)^{\frac{\min(s,d)}{d}}\right)
\]
\end{cor}
The corollary still holds if $N_{1}$ and $N_{2}$ are linearly reated
and if $k_{1}$ and $k_{2}$ are linearly related. An ensemble of
estimators is formed by choosing different neighborhood sizes by choosing
different values of $k$. Choose $\mathcal{L}=\left\{ l_{1},\dots,l_{L}\right\} $
to be real positive numbers that index $h(l_{i})$. Define $w:=\left\{ w\left(l_{1}\right),\dots,w\left(l_{L}\right)\right\} $
and $\hat{\mathbf{G}}_{w}:=\sum_{l\in\mathcal{L}}w(l)\hat{\mathbf{G}}_{k(l)}$.
The weights can be used to decrease the bias as before.

An ensemble of estimators is formed by choosing different neighborhood
sizes by choosing different values of $k$. Choose $\mathcal{L}=\left\{ l_{1},\dots,l_{L}\right\} $
to be real positive numbers that index $k(l_{i})$. Define $w:=\left\{ w\left(l_{1}\right),\dots,w\left(l_{L}\right)\right\} $
and $\hat{\mathbf{G}}_{w}:=\sum_{l\in\mathcal{L}}w(l)\hat{\mathbf{G}}_{k(l)}$.
The weights can be used to decrease the bias as before. Consider the
following assumptions on the ensemble of estimators $\left\{ \hat{\mathbf{G}}_{k(l)}\right\} _{l\in\mathcal{L}}$
\cite{moon2016isit}:
\begin{itemize}
\item $\mathcal{C}.1$ The bias is expressible as 
\[
\bias\left[\hat{\mathbf{G}}_{k(l)}\right]=\sum_{i\in J}c_{i}\psi_{i}(l)\phi_{i,d}(N)+O\left(\frac{1}{\sqrt{N}}\right),
\]
where $c_{i}$ are constants depending on the underlying density and
are independent of $N$ and $l$, $J=\left\{ i_{1},\dots,i_{I}\right\} $
is a finite index set with $I<L$, and $\psi_{i}(l)$ are basis functions
depending only on the parameter $l$ and not on the sample size $N$.
\item $\mathcal{C}.2$ The variance is expressible as 
\[
\var\left[\hat{\mathbf{G}}_{k(l)}\right]=c_{v}\left(\frac{1}{N}\right)+o\left(\frac{1}{N}\right).
\]
\end{itemize}
\begin{thm}
\cite{moon2016isit} Assume conditions $\mathcal{C}.1$ and $\mathcal{C}.2$
hold for the ensemble of estimators $\left\{ \hat{\mathbf{G}}_{k(l)}\right\} _{l\in\mathcal{L}}$.
Then there exists a weight vector $w_{0}$ such that the MSE of the
weighted ensemble estimator attains the parametric rate of convergence:
\[
\bE\left[\left(\hat{\mathbf{G}}_{w_{0}}-G\left(f_{1},f_{2}\right)\right)^{2}\right]=O\left(\frac{1}{N}\right).
\]
The weight vector $w_{0}$ is the solution to the following offline
convex optimization problem:
\begin{equation}
\begin{array}{rl}
\min_{w} & ||w||_{2}\\
subject\,to & \sum_{l\in\mathcal{L}}w(l)=1,\\
 & \gamma_{w}(i)=\sum_{l\in\mathcal{L}}w(l)\psi_{i}(l)=0,\,i\in J.
\end{array}\label{eq:optim}
\end{equation}
\end{thm}
To achieve the parametric rate $O(1/N)$ in MSE convergence, it is
not necessary that $\gamma_{w}(i)=0$, $i\in J$. The following convex
optimization is also sufficient \cite{moon2016isit,wisler2017direct}:

\begin{equation}
\begin{array}{rl}
\min_{w} & \epsilon\\
subject\,to & \sum_{\ell\in\bar{\ell}}w(\ell)=1,\\
 & \left|\gamma_{w}(i)N^{\frac{1}{2}}\phi_{i,d}(N)\right|\leq\epsilon,\,i\in\{1,\dots,J\},\\
 & ||w||_{2}^{2}\leq\eta\epsilon,
\end{array}\label{eq:relaxed}
\end{equation}
where the parameter $\eta$ is chosen to achieve a trade-off between
bias and variance. 

We now aply this theory to the plug-in $k$-nn estimators. For general
$g$, let $k(l)=l\sqrt{N}$. From Theorem~\ref{thm:biasKNN}, we
have $\psi_{i}(l)=l^{i/d}$ for $i=1,\dots,d$. If $s\geq d$, then
we have a $O\left(\frac{1}{l\sqrt{N}}\right)$. We also include the
function $\psi_{d+1}(l)=l^{-1}$. The bias of the resulting base estimator
satisfies condition $\mathcal{C}.1$ with $\phi_{i,d}(N)=N^{-i/(2d)}$
for $i=1,\dots,d$ and $\phi_{i,d+1}(N)=N^{-1/2}$. The variance also
satisfies condition $\mathcal{C}.2$. The optimal weight $w_{0}$
is found using (\ref{eq:relaxed}) to obtain a plug-in divergence
functional estimator $\hat{\mathbf{G}}_{w_{0},1}$ with an MSE convergence
rate of $O\left(\frac{1}{N}\right)$ as long as $s\geq d$. Otherwise,
if $s<d$ we can only guarantee the MSE rate up to $O\left(\frac{1}{N^{s/d}}\right)$.
We refer to this estimator as the ODin1 $k$-nn estimator.

We can define another weighted ensemble estimator that achieves the
parametric rate under less strict assumptions on the smoothness of
the densities if the functional $g$ satisfies the assumption required
for (\ref{eq:bias2KNN}). Let $\delta>0$ and $k(l)=lN^{\delta}$.
From Theorem~\ref{thm:biasKNN}, the bias has terms proportional
to $l^{j-\frac{q}{2}}N^{-\frac{(1-\delta)j}{d}-\frac{q\delta}{2}}$
where $j,q\geq0$ and $j+\frac{q}{2}>\frac{1}{2}$. Let $\phi_{j,q,d}(N)=N^{-\frac{(1-\delta)j}{d}-\frac{q\delta}{2}}$
and $\psi_{j,q}(l)=l^{j-\frac{q}{2}}$. Let 
\begin{eqnarray*}
J & = & \left\{ \left\{ j,q\right\} :0<\frac{(1-\delta)j}{d}+\frac{q\delta}{2}<\frac{1}{2},\,q\in\{0,1,2,\dots,\nu\},j\in\{0,1,2,\dots,r\},\,j+\frac{q}{2}>\frac{1}{2}\right\} 
\end{eqnarray*}
Then the bias of the resulting base estimator satisfies condition
$\mathcal{C}.1$ and the variance satisfies condition $\mathcal{C}.2$.
If $L>|J|$, then the optimal weight can be found using (\ref{eq:relaxed}).
The resulting weighted ensemble estimator $\hat{\mathbf{G}}_{w_{0},2}$
achieves the parametric convergence rate if $\nu\geq1/\delta$ and
if $s\geq\frac{d}{2(1-\delta)}$. Otherwise, if $s<d/(2(1-\delta))$
we can only guarantee the MSE rate up to $O\left(\frac{1}{N^{\frac{2(1-\delta)s}{d}}}\right)$.
We refer to this estimator as the ODin2 $k$-nn estimator.

The parametric rate can be achieved with $\hat{\mathbf{G}}_{w_{0},2}$
under less strict assumptions on the smoothness of the densities than
those required for $\hat{\mathbf{G}}_{w_{0},1}$. Since $\delta>0$
can be arbitrary, it is theoretically possible to construct an estimator
that achieves the parametric rate as long as $s>d/2$. However, $\hat{\mathbf{G}}_{w_{0},2}$
requires more parameters to implement the weighted ensemble estimator
than $\hat{\mathbf{G}}_{w_{0},1}$ which may have an effect on the
variance.

\subsection{Central Limit Theorem}

The following theorem shows that the appropriately normalized ensemble
estimator $\hat{\mathbf{G}}_{w}$ converges in distribution to a normal
random variable, which enables us to perform hypothesis testing on
the divergence functional. The proof uses a lemma modified from~\cite{sricharan2012estimation}
that gives sufficient conditions on an interchangeable process for
a central limit theorem. The details are given in Appendix~\ref{sec:cltProofKNN}.
\begin{thm}
 \label{thm:cltKNN}Assume that the mixed derivatives of $g$ of order
$2$ are bounded and $k(l)\rightarrow\infty$ as $N\rightarrow\infty$
for each $l\in\mathcal{L}$. Then for fixed $L$, and if $\mathbf{S}$
is a standard normal random variable, 
\[
\Pr\left(\left(\hat{\mathbf{G}}_{w}-\bE\left[\hat{\mathbf{G}}_{w}\right]\right)/\sqrt{\var\left[\hat{\mathbf{G}}_{w}\right]}\leq t\right)\rightarrow\Pr\left(\mathbf{S}\leq t\right).
\]
\end{thm}

\section{Numerical Validation}

\label{sec:experiments}We validate our theory on the MSE convergence
rates by estimating the Rényi-$\alpha$ divergence integral between
two truncated multivariate Gaussian distributions with varying dimension
and sample sizes. The densities have means $\bar{\mu}_{1}=0.7*\bar{1}_{d}$,
$\bar{\mu}_{2}=0.3*\bar{1}_{d}$ and covariance matrices $0.1*I_{d}$
where $\bar{1}_{d}$ is a $d$-dimensional vector of ones, and $I_{d}$
is a $d\times d$ identity matrix. We used $\alpha=0.5$ and restricted
the Gaussians to the unit cube.

\begin{figure}
\centering \includegraphics[width=0.5\textwidth]{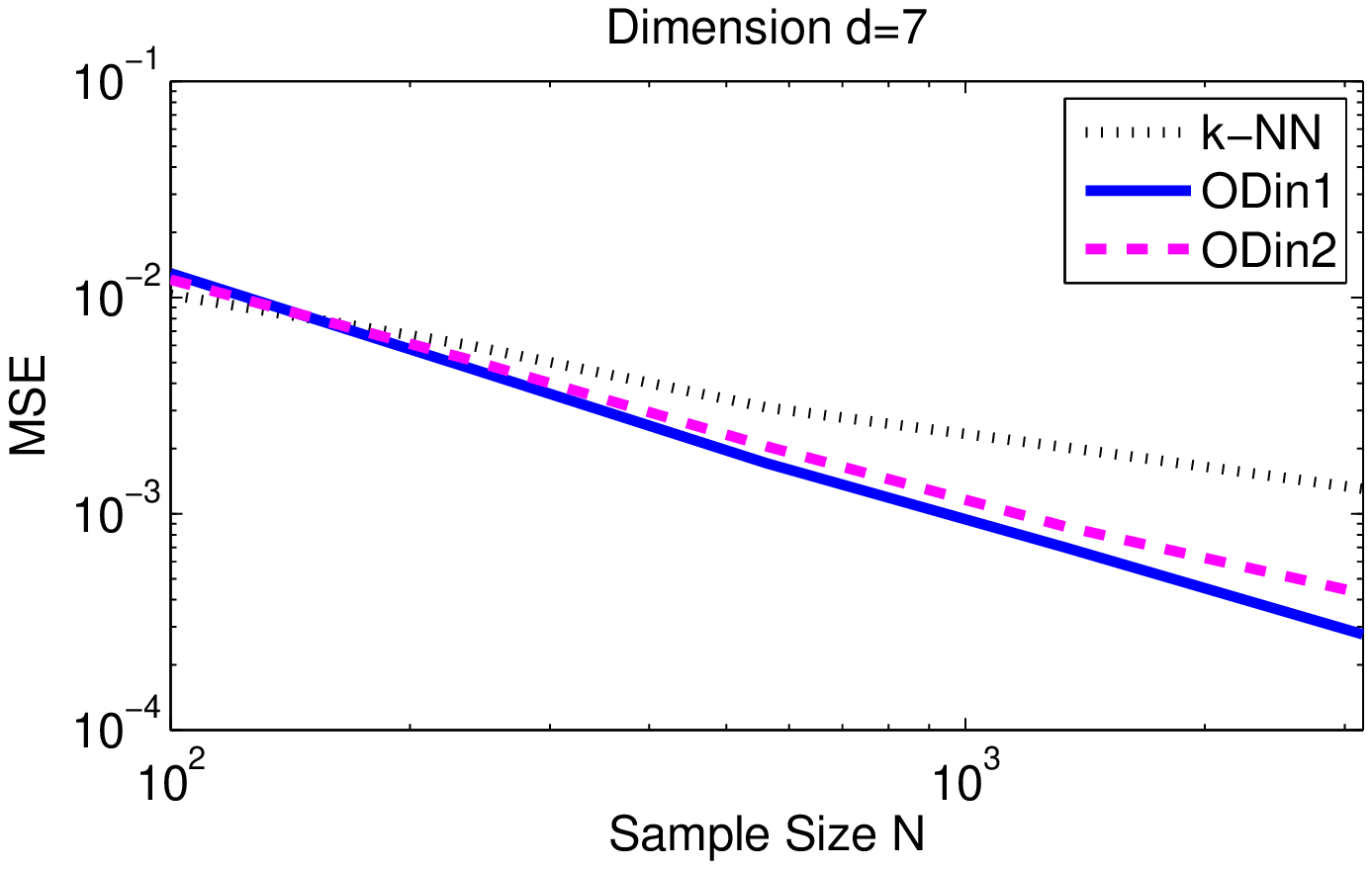}\includegraphics[width=0.5\textwidth]{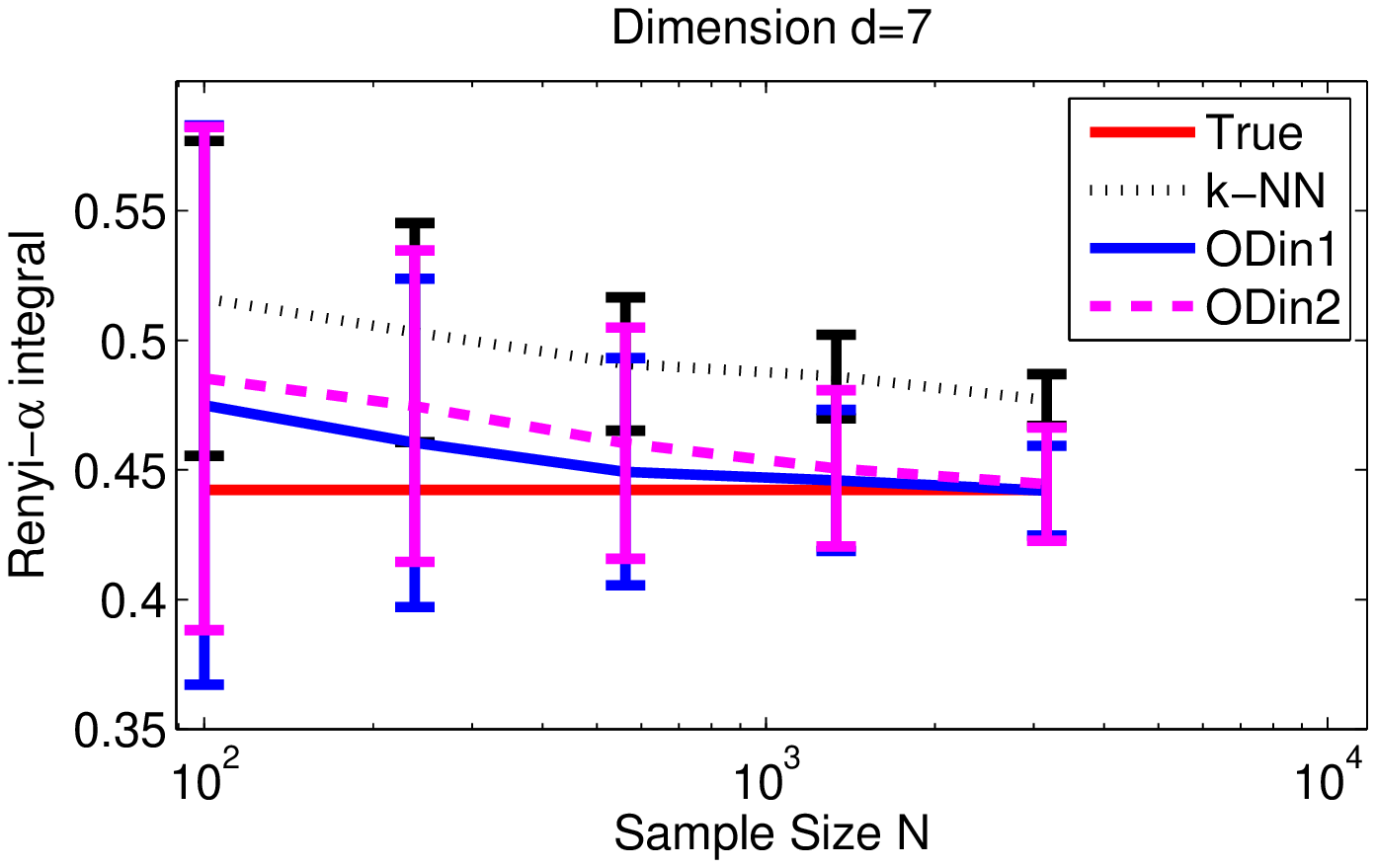}
\caption{\label{fig:mseKNN} (Left) Log-log plot of MSE of the $k$-nn plug-in
estimator (``k-NN\char`\"{}) and the two proposed optimally weighted
estimators (ODin1 and ODin2) for $d=7$. (Right) Plot of the average
value of the same estimators with standard error bars compared to
the true value being estimated. The proposed weighted ensemble estimators
outperform the plug-in estimator.}
\end{figure}

The left plot in Fig.~\ref{fig:mseKNN} shows the MSE (200 trials)
of the standard plug-in $k$-nn estimator where $k=\sqrt{N}$ and
the two proposed optimally weighted estimators ODin1 and ODin2. We
show the case where $d=7$ and the sample size varies. For the ODin1
estimator, we chose $\mathcal{L}$ to be linearly spaced between $0.3$
and $3$ with $L=50$. For the ODin2 estimator, we chose the minimum
value of $\mathcal{L}$ to be 1.4 and then chose the next 24 values
for $k$ (i.e. $L=25$). Both ODin1 and ODin2 outperform both plug-in
estimators which validates our theory.

\section{Conclusion}

In this chapter, we derived convergence rates for a $k$-nearest neighbor
plug-in estimator of divergence functionals. We applied the generalized
theory of optimally weighted ensemble estimation derived previously
to derive an estimator that achieves the parametric rate when the
densities belong to the Hölder smoothness class with smoothness parameter
greater than $d/2$. The convergence rates we derive apply when the
densities have support $[0,1]^{d}$ although the estimators do not
require knowledge of the support. We also derived the asymptotic distribution
of the estimator.

\bibliographystyle{IEEEtran}
\bibliography{References}

\appendices{}

\section{The Boundary Condition}

\label{sec:boundary_sphere}In this section, we prove a result on
the boundary of the densities' support set that will be necessary
to derive the bias expressions in Theorem~\ref{thm:biasKNN}. Consider
a uniform circular kernel $K(x)$ with $K(x)=1$ for all $x$ s.t.
$||x||_{2}\leq1$. We also consider the family of probability densities
with rectangular support $\mathcal{S}=[-1,1]^{d}$. We show that the
following smoothness condition holds: for any polynomial $p_{x}(u):\mathbb{R}^{d}\rightarrow\mathbb{R}$
of degree $q\leq r=\left\lfloor s\right\rfloor $ with coefficients
that are $r-q$ times differentiable wrt $x$, 
\begin{equation}
\int_{x\in\mathcal{S}}\left(\int_{u:||u||_{2}\leq1,x+uh\notin\mathcal{S}}p_{x}(u)du\right)^{t}dx=v_{t}(h),\label{eq:boundary}
\end{equation}
 where $v_{t}(h)$ has the expansion 
\[
v_{t}(h)=\sum_{i=1}^{r-q}e_{i,q,t}h^{i}+o(h^{r-q}).
\]

Note that the inner integral forces the $x$ terms to be boundary
points through the constraint $x+uh\notin\mathcal{S}$. Note also
that this proof is more difficult than for the uniform rectangular
kernel in \cite{moon2016arxiv} since in that case, the kernel aligns
better with the boundary.

\subsection{Single Coordinate Boundary Point}

\label{sub:single_euc}We begin by focusing on points $x$ that are
boundary points due to a single coordinate $x_{i}$ s.t. $x_{i}+u_{i}h\notin\mathcal{S}$.
Without loss of generality, assume that $x_{i}+u_{i}h>1$. We focus
first on the inner integral in (\ref{eq:boundary}). We will use the
following lemma: 
\begin{lem}
\label{lem:poly_sphere}Let $D_{d}(\rho)$ be a $d$-sphere with radius
$d$ and let $\sum_{i=1}^{d}n_{i}=q$. Then 
\[
\int_{D_{d}(r)}u_{1}^{n_{1}}u_{2}^{n_{2}}\dots u_{d}^{n_{d}}du_{1}\dots du_{d}=C\rho^{d+q},
\]
 where $C$ is a constant that depends on the $n_{i}$s and $d$.
\end{lem}
\begin{IEEEproof}
We convert to $d$-dimensional spherical coordinates to handle the
integration. Let $r$ be the distance of a point $u$ from the origin.
We nave $d-1$ angular coordinates $\phi_{i}$ where $\phi_{d-1}$
ranges from $0$ to $2\pi$ and all other $\phi_{i}$ range from $0$
to $\pi$. The conversion from the spherical coordinates to Cartesian
coordinates is then 
\begin{eqnarray*}
u_{1} & = & r\cos\left(\phi_{1}\right)\\
u_{2} & = & r\sin\left(\phi_{1}\right)\cos\left(\phi_{2}\right)\\
u_{3} & = & r\sin\left(\phi_{1}\right)\sin\left(\phi_{2}\right)\cos\left(\phi_{3}\right)\\
\vdots\\
u_{d-1} & = & r\sin\left(\phi_{1}\right)\cdots\sin\left(\phi_{d-2}\right)\cos\left(\phi_{d-1}\right)\\
u_{d} & = & r\sin\left(\phi_{1}\right)\cdots\sin\left(\phi_{d-2}\right)\sin\left(\phi_{d-1}\right).
\end{eqnarray*}
 The spherical volume element is then 
\[
r^{d-1}\sin^{d-2}\left(\phi_{1}\right)\sin^{d-3}\left(\phi_{1}\right)\cdots\sin\left(\phi_{d-1}\right)dr\,d\phi_{1}\,d\phi_{2}\cdots d\phi_{d-1}.
\]
 Combining these results gives 
\begin{eqnarray*}
\lefteqn{\int_{D_{d}(r)}u_{1}^{n_{1}}u_{2}^{n_{2}}\dots u_{d}^{n_{d}}du_{1}\dots du_{d}}\\
 & = & \int_{0}^{\rho}\int_{o}^{2\pi}\int_{0}^{\pi}\cdots\int_{0}^{\pi}r^{q+d-1}\left[\sin^{q-n_{1}+d-2}\left(\phi_{1}\right)\sin^{q-n_{1}-n_{d}+d-3}\left(\phi_{2}\right)\cdots\right.\\
 &  & \left.\sin^{n_{d}+n_{d-1}+1}\left(\phi_{d-2}\right)\sin^{n_{d}}\left(\phi_{d-1}\right)\right]\left[\cos^{n_{1}}\left(\phi_{1}\right)\cdots\cos^{n_{d}}\left(\phi_{d-1}\right)\right]d\phi_{1}\cdots d\phi_{d-1}dr\\
 & = & C\rho^{q+d}.
\end{eqnarray*}
\end{IEEEproof}
The region of integration for the inner integral in (\ref{eq:boundary})
corresponds to a hyperspherical cap with radius $1$ and height of
$\frac{1-x_{i}}{h}$. The inner integral can be calculated using an
approach similar to that used in~\cite{li2011spherical_cap} to calculate
the volume of a hyperspherical cap. It is obtained by integrating
the polynomial $p_{x}(u)$ over a $d-1$-sphere with radius $\sin\theta$
and height element $d\cos\theta$. This is done using Lemma~\ref{lem:poly_sphere}.
We then integrate over $\theta$ which has a range of $0$ to $\phi=\cos^{-1}\left(\frac{1-x_{i}}{h}\right).$
Thus we have 
\begin{eqnarray}
\int_{u:||u||_{2}\leq1,x+uh\notin\mathcal{S}}p_{x}(u)du & = & \sum_{m=0}^{q}\tilde{p}_{m}(x)\int_{0}^{\phi}\sin^{m+d-1}(\theta)\sin\theta u_{d}^{m}d\theta\nonumber \\
 & = & \sum_{m=0}^{q}\tilde{p}_{m}(x)\int_{0}^{\phi}\sin^{m+d}(\theta)\cos^{m}\theta d\theta.\label{eq:poly1}
\end{eqnarray}
From standard integral tables, we get that for $n\geq2$ and $m\geq0$
\begin{equation}
\int_{0}^{\phi}\sin^{n}\theta\cos^{m}\theta d\theta=-\frac{\sin^{n-1}\phi\cos^{m+1}\phi}{n+m}+\frac{n-1}{n+m}\int_{0}^{\phi}\sin^{n-2}\theta\cos^{m}\theta d\theta.\label{eq:sincos}
\end{equation}
If $n=1$, then we get 
\[
\int_{0}^{\phi}\sin\theta\cos^{m}\theta d\theta=\frac{1}{m+1}-\frac{\cos^{m+1}\phi}{m+1}.
\]
Since $\phi=\cos^{-1}\left(\frac{1-x_{i}}{h}\right)$, we have 
\begin{eqnarray*}
\cos\phi & = & \frac{1-x_{i}}{h},\\
\sin\phi & = & \sqrt{1-\left(\frac{1-x_{i}}{h}\right)^{2}}.
\end{eqnarray*}
Therefore, if $n$ is odd, we obtain 
\begin{equation}
\int_{0}^{\phi}\sin^{n}\theta\cos^{m}\theta d\theta=\sum_{\ell=0}^{(n-1)/2}c_{\ell}\left(\sqrt{1-\left(\frac{1-x_{i}}{h}\right)^{2}}\right)^{2\ell}\left(\frac{1-x_{i}}{h}\right)^{m+1}+c,\label{eq:nodd}
\end{equation}
where the constants depend on $m$ and $n$.

If $n$ is even and $m>0$, then the final term in the recursion in
(\ref{eq:sincos}) reduces to 
\[
\int_{0}^{\phi}\cos^{m}\theta d\theta=\frac{\cos^{m-1}\phi\sin\phi}{m}+\frac{m-1}{m}\int_{0}^{\phi}\cos^{m-2}\theta d\theta.
\]
If $m=2$, then 
\begin{eqnarray*}
\int_{0}^{\phi}\cos^{2}\theta d\theta & = & \frac{\phi}{2}+\frac{1}{4}\sin(2\phi)\\
 & = & \frac{\phi}{2}+\frac{1}{2}\sin\phi\cos\phi.
\end{eqnarray*}
Therefore, if $n$ and $m$ are both even, then this gives 
\begin{eqnarray}
\int_{0}^{\phi}\sin^{n}\theta\cos^{m}\theta d\theta & = & \sum_{\ell=0}^{(n-2)/2}c_{\ell}^{'}\left(\sqrt{1-\left(\frac{1-x_{i}}{h}\right)^{2}}\right)^{2\ell+1}\left(\frac{1-x_{i}}{h}\right)^{m+1}+c^{'}\cos^{-1}\left(\frac{1-x_{i}}{h}\right)\nonumber \\
 &  & +\sum_{\ell=0}^{(m-2)/2}c_{\ell}^{''}\left(\sqrt{1-\left(\frac{1-x_{i}}{h}\right)^{2}}\right)\left(\frac{1-x_{i}}{h}\right)^{2\ell+1}.\label{eq:nm_even}
\end{eqnarray}
On the other hand, if $n$ is even and $m$ is odd, we get 
\begin{eqnarray}
\int_{0}^{\phi}\sin^{n}\theta\cos^{m}\theta d\theta & = & \sum_{\ell=0}^{(n-2)/2}c_{\ell}^{'''}\left(\sqrt{1-\left(\frac{1-x_{i}}{h}\right)^{2}}\right)^{2\ell+1}\left(\frac{1-x_{i}}{h}\right)^{m+1}\nonumber \\
 &  & +\sum_{\ell=0}^{(m-1)/2}c_{\ell}^{''''}\left(\sqrt{1-\left(\frac{1-x_{i}}{h}\right)^{2}}\right)\left(\frac{1-x_{i}}{h}\right)^{2\ell}.\label{eq:neven_modd}
\end{eqnarray}

If $d$ is odd, then combining (\ref{eq:nodd}) and (\ref{eq:neven_modd})
with (\ref{eq:poly1}) gives 
\begin{align}
\int_{u:||u||_{2}\leq1,x+uh\notin\mathcal{S}}p_{x}(u)du & = & \sum_{m=0}^{q}\sum_{\ell=0}^{d+q}p_{m,\ell}(x)\left(\sqrt{1-\left(\frac{1-x_{i}}{h}\right)^{2}}\right)^{\ell}\left(\frac{1-x_{i}}{h}\right)^{m},\label{eq:polyodd}
\end{align}
where the coefficients $p_{m,\ell}(x)$ are $r-q$ times differentiable
wrt $x$. Similarly, if $d$ is even, then 
\begin{eqnarray}
\int_{u:||u||_{2}\leq1,x+uh\notin\mathcal{S}}p_{x}(u)du & = & \sum_{m=0}^{q}\sum_{\ell=0}^{d+q}p_{m,\ell}^{'}(x)\left(\sqrt{1-\left(\frac{1-x_{i}}{h}\right)^{2}}\right)^{\ell}\left(\frac{1-x_{i}}{h}\right)^{m}\nonumber \\
 &  & +p^{'}(x)\cos^{-1}\left(\frac{1-x_{i}}{h}\right),\label{eq:polyeven}
\end{eqnarray}
where again the coefficients $p_{m,\ell}^{'}(x)$ and $p^{'}(x)$
are $r-q$ times differentiable wrt $x$. Raising (\ref{eq:polyodd})
and (\ref{eq:polyeven}) to the power of $t$ gives respective expressions
of the form 
\begin{equation}
\sum_{m=0}^{qt}\sum_{\ell=0}^{(d+q)t}\check{p}_{m,\ell}(x)\left(\sqrt{1-\left(\frac{1-x_{i}}{h}\right)^{2}}\right)^{\ell}\left(\frac{1-x_{i}}{h}\right)^{m},\label{eq:polyodd2}
\end{equation}
\begin{equation}
\sum_{m=0}^{qt}\sum_{\ell=0}^{(d+q)t}\sum_{n=0}^{t}\check{p}_{m,\ell,n}(x)\left(\sqrt{1-\left(\frac{1-x_{i}}{h}\right)^{2}}\right)^{\ell}\left(\frac{1-x_{i}}{h}\right)^{m}\left(\cos^{-1}\left(\frac{1-x_{i}}{h}\right)\right)^{n},\label{eq:polyeven2}
\end{equation}
where the coefficients $\check{p}_{m,\ell}(x)$ and $\check{p}_{m,\ell,n}(x)$
are all $r-q$ times differentiable wrt $x$. Integrating (\ref{eq:polyodd2})
and (\ref{eq:polyeven2}) over all the coordinates in $x$ except
for $x_{i}$ affects only the $\check{p}_{m,\ell}(x)$ and $\check{p}_{m,\ell,n}(x)$
coefficients, resulting in respective expressions of the form 
\begin{equation}
\sum_{m=0}^{qt}\sum_{\ell=0}^{(d+q)t}\bar{p}_{m,\ell}(x_{i})\left(\sqrt{1-\left(\frac{1-x_{i}}{h}\right)^{2}}\right)^{\ell}\left(\frac{1-x_{i}}{h}\right)^{m},\label{eq:polyodd3}
\end{equation}
\begin{equation}
\sum_{m=0}^{qt}\sum_{\ell=0}^{(d+q)t}\sum_{n=0}^{t}\bar{p}_{m,\ell,n}(x_{i})\left(\sqrt{1-\left(\frac{1-x_{i}}{h}\right)^{2}}\right)^{\ell}\left(\frac{1-x_{i}}{h}\right)^{m}\left(\cos^{-1}\left(\frac{1-x_{i}}{h}\right)\right)^{n}.\label{eq:polyeven3}
\end{equation}
The coefficients $\bar{p}_{m,\ell}(x_{i})$ and $\bar{p}_{m,\ell,n}(x_{i})$
are $r-q$ times differentiable wrt $x_{i}$. Since the other coordinates
of $x$ other than $x_{i}$ are far away from the boundary, the coefficients
are independent of $h$. For the integral wrt $x_{i}$ of (\ref{eq:polyodd3}),
taking a Taylor series expansion of $\bar{p}_{m,\ell}(x_{i})$ around
$x_{i}=1$ yields terms of the form 
\begin{eqnarray*}
\int_{1-h}^{1}\left(\sqrt{1-\left(\frac{1-x_{i}}{h}\right)^{2}}\right)^{\ell}\left(\frac{1-x_{i}}{h}\right)^{m+j}h^{j}dx_{i} & = & h^{j+1}\int_{0}^{1}\left(1-y_{i}\right)^{\frac{\ell}{2}}y_{i}^{\frac{m+j-1}{2}}dy_{i}\\
 & = & h^{j+1}B\left(\frac{\ell+2}{2},\frac{m+j+1}{2}\right),
\end{eqnarray*}
where $0\leq j\leq r-q$, $0\leq\ell\leq(d+q)t$, $0\leq m\leq qt$,
and $B(x,y)$ is the beta function. Note that the first step uses
the substitution of $y_{i}=\left(\frac{1-x_{i}}{h}\right)^{2}$.

If $d$ is even (i.e. (\ref{eq:polyeven3})), a simple closed-form
expression is not easy to obtain due to the $\cos^{-1}\left(\frac{1-x_{i}}{h}\right)$
terms. However, by similarly applying a Taylor series expansion to
$\bar{p}_{m,\ell,n}(x_{i})$ and substituting $y_{i}=\frac{1-x_{i}}{h}$
gives terms of the form of 
\begin{eqnarray*}
\lefteqn{\int_{1-h}^{1}\left(\sqrt{1-\left(\frac{1-x_{i}}{h}\right)^{2}}\right)^{\ell}\left(\frac{1-x_{i}}{h}\right)^{m+j}\left(\cos^{-1}\left(\frac{1-x_{i}}{h}\right)\right)^{n}h^{j}dx_{i}}\\
 & = & h^{j+1}\int_{0}^{1}\left(1-y_{i}^{2}\right)^{\frac{\ell}{2}}y_{i}^{m+j}\left(\cos^{-1}y_{i}\right)^{n}dy_{i}\\
 & = & h^{j+1}c_{\ell,m,j,n},
\end{eqnarray*}
for $0\leq j\leq r-q$, $0\leq\ell\leq(d+q)t$, $0\leq m\leq qt$,
and $0\leq n\leq t$. Combining terms results in the expansion $v_{t}(h)=\sum_{i=1}^{r-q}e_{i,q,t}h^{i}+o(h^{r-q})$.

\subsection{Multiple Coordinate Boundary Point}

\label{sec:bound_sphere_mult}The case where multiple coordinates
of the point $x$ are near the boundary is a fairly straightforward
extension of the single boundary point case. Consider the case where
2 of the coordinates are near the boundary, e.g., $x_{1}$ and $x_{2}$
with $x_{1}+u_{1}h>1$ and $x_{2}+u_{2}h>1$. The region of integration
for the inner integral can be decomposed into two parts: a hyperspherical
cap wrt $x_{1}$ and the remaining area (denoted, respectively, as
$A_{1}$ and $A_{2}$). The remaining area $A_{2}$ can be decomposed
further into two other areas: a hyperspherical cap wrt $x_{2}$ (denoted
$B_{1}$) and a height chosen s.t. $B_{1}$ just intersects $A_{1}$
on their boundaries. Integrating over the remainder of $A_{2}$ is
achieved by integrating along $x_{2}$ over $d-1$-dimensional hyperspherical
caps from the boundary of $B_{1}$ to the boundary of $A_{2}$. Thus
integrating over these regions yields an expression similar to (\ref{eq:poly1}).
Following a similar procedure will then yield the result. 

\section{Proof of Theorem~\ref{thm:biasKNN} (Bias)}

\label{sec:biasProofKNN}In this section, we prove the bias results
in Thm. \ref{thm:biasKNN}. The bias of the base $k$-nn plug-in estimator
$\ghat$ can be expressed as 
\begin{eqnarray}
\bias\left[\ghat\right] & = & \bE\left[g\left(\fh 1(\mathbf{Z}),\fh 2(\mathbf{Z})\right)-g\left(f_{1}(\mathbf{Z}),f_{2}(\mathbf{Z})\right)\right]\nonumber \\
 & = & \bE\left[g\left(\fh 1(\mathbf{Z}),\fh 2(\mathbf{Z})\right)-g\left(\bE_{\mathbf{Z},\dist 1(\mathbf{Z})}\fh 1(\mathbf{Z}),\bE_{\mathbf{Z},\dist 2(\mathbf{Z})}\fh 2(\mathbf{Z})\right)\right]\nonumber \\
 &  & +\bE\left[g\left(\bE_{\mathbf{Z},\dist 1(\mathbf{Z})}\fh 1(\mathbf{Z}),\bE_{\mathbf{Z},\dist 2(\mathbf{Z})}\fh 2(\mathbf{Z})\right)-g\left(f_{1}(\mathbf{Z}),f_{2}(\mathbf{Z})\right)\right],\label{eq:gsplit}
\end{eqnarray}
where $\mathbf{Z}$ is drawn from $f_{2}$ and $\dist i(\mathbf{Z})$
is the $k_{i}$th nearest neighbor distance of $\mathbf{Z}$ in the
respective samples. For notational simplicity, let $\dist i(\mathbf{Z})=\dist i$.
The $k$-nn density estimator can be viewed as a kernel density estimator.
Let $K$ be the uniform kernel on the unit ball. That is, 
\[
K(x)=\begin{cases}
\frac{1}{c_{d}}, & ||x||<1\\
0, & \text{otherwise},
\end{cases}
\]
where $c_{d}$ is the volume of the unit ball in $\mathbb{R}^{d}$.
Then we have that 
\begin{eqnarray*}
\fh 1(\mathbf{Z}) & = & \frac{1}{N_{1}\dist 1^{d}}\sum_{i=1}^{N_{1}}K\left(\frac{\mathbf{Z}-\mathbf{Y}_{i}}{\dist 1}\right),\\
\fh 2(\mathbf{Z}) & = & \frac{1}{N_{2}\dist 2^{d}}\sum_{i=1}^{N_{2}}K\left(\frac{\mathbf{Z}-\mathbf{X}_{i}}{\dist 2}\right).
\end{eqnarray*}
The fact that the $k$-nn distances are random requires extra care.
However, we can condition on these distances with these representations
which enables us to use some of the same tools as in the KDE approach
\cite{moon2016arxiv}. Define 
\begin{eqnarray*}
S_{k_{i}}(\mathbf{Z}) & = & \left\{ X\in\mathbb{R}^{d}:\left\Vert X-\mathbf{Z}\right\Vert <\dist i\right\} ,\\
\implies\Pr\left(S_{k_{i}}(\mathbf{Z})\right) & = & \int_{S_{k_{i}}(\mathbf{Z})}f_{i}(x)dx.
\end{eqnarray*}
Note that from \cite{mack1979multivariate}, we have that 
\begin{equation}
\bE_{\mathbf{Z},\dist i}\fh i(\mathbf{Z})=\frac{k_{i}-1}{N_{i}}\frac{1}{\dist i^{d}}\frac{1}{\Pr\left(S_{k_{i}}(\mathbf{Z})\right)}\int_{S_{k_{i}}(\mathbf{Z})}K\left(\frac{\mathbf{Z}-x}{\dist i}\right)f_{i}(x)dx.\label{eq:conditional_knn}
\end{equation}

The Taylor series expansion of $g\left(\bE_{\mathbf{Z},\dist 1}\fh 1(\mathbf{Z}),\bE_{\mathbf{Z},\dist 2}\fh 2(\mathbf{Z})\right)$
around $f_{1}(\mathbf{Z})$ and $f_{2}(\mathbf{Z})$ is 
\begin{align}
g\left(\bE_{\mathbf{Z},\dist 1}\fh 1(\mathbf{Z}),\bE_{\mathbf{Z},\dist 2}\fh 2(\mathbf{Z})\right) & = & \sum_{i=0}^{\infty}\sum_{j=0}^{\infty}\left(\left.\frac{\partial^{i+j}g(x,y)}{\partial x^{i}\partial y^{j}}\right|_{\substack{x=f_{1}(\mathbf{Z})\\
y=f_{2}(\mathbf{Z})
}
}\right)\nonumber \\
 &  & \times\frac{\bias_{\mathbf{Z},\dist 1}^{i}\left[\fh 1(\mathbf{Z})\right]\bias_{\mathbf{Z},\dist 2}^{j}\left[\fh 2(\mathbf{Z})\right]}{i!j!},\label{eq:bias_term}
\end{align}
where $\bias_{\mathbf{Z},\dist i}^{j}\left[\fh i(\mathbf{Z})\right]=\left(\bE_{\mathbf{Z},\dist i}\fh i(\mathbf{Z})-f_{i}(\mathbf{Z})\right)^{j}$.
We thus require an expression for $\bias_{\mathbf{Z},\dist i}\left[\fh i(\mathbf{Z})\right]$.
Since we are conditioning on $\dist i$, we can consider separately
the cases when $\mathbf{Z}$ is in the interior of the support $\mathcal{S}$
or when $\mathbf{Z}$ is near the boundary of the support. As before,
A point $X\in\mathcal{S}$ is defined to be in the interior of$\mathcal{S}$
if for all $Y\notin\mathcal{S}$, $K\left(\frac{X-Y}{h_{i}}\right)=0$.
A point$X\in\mathcal{S}$ is near the boundary of the support if it
is not in the interior. Denote the region in the interior and near
the boundary wrt $\dist i$ as $\mathcal{S}_{I_{i}}$ and $\mathcal{S}_{B_{i}}$,
respectively. Recall that we assume that $\mathcal{S}=[0,1]^{d}$,
the unit cube.

Consider now $\int_{S_{k_{i}}(\mathbf{Z})}K\left(\frac{\mathbf{Z}-x}{\dist i}\right)f_{i}(x)dx$.
Substituting $u=\frac{x-\mathbf{Z}}{\dist i}$ and then taking a Taylor
series expansion of $f_{i}$ using multi-index notation gives 
\begin{eqnarray*}
\int_{S_{k_{i}}(\mathbf{Z})}K\left(\frac{x-\mathbf{Z}}{\dist i}\right)f_{i}(x)dx & = & \dist i^{d}\int_{||u||<1}K\left(u\right)f_{i}(\mathbf{Z}+u\dist i)du\\
 & = & \sum_{|\alpha|\leq\left\lfloor s\right\rfloor }\frac{D^{\alpha}f_{i}(\mathbf{Z})}{\alpha!}\dist i^{d+|\alpha|}\int_{u:\mathbf{Z}+u\dist i\in\mathcal{S}}u^{\alpha}K(u)du,+O\left(\dist i^{d+s}\right)
\end{eqnarray*}
\begin{align}
\implies\bE_{\mathbf{Z},\dist i}\fh i(\mathbf{Z}) & = & \frac{k_{i}-1}{N_{i}}\frac{1}{\Pr\left(S_{k_{i}}(\mathbf{Z})\right)}\nonumber \\
 &  & \times\left(\sum_{|\alpha|\leq\left\lfloor s\right\rfloor }\frac{D^{\alpha}f_{i}(\mathbf{Z})}{\alpha!}\dist i^{|\alpha|}\int_{u:\mathbf{Z}+u\dist i\in\mathcal{S}}u^{\alpha}K(u)du+O\left(\dist i^{s}\right)\right).\label{eq:conditional_knn2}
\end{align}

\begin{lem}
\label{lem:exp_bias}Let $\gamma(x,y)$ be an arbitrary function satisfying
$\sup_{x,y}|\gamma(x,y)|<\infty$. Let $\mathcal{S}=[0,1]^{d}$ and
let $f_{1},f_{2}\in\Sigma(s,L)$. Let $\mathbf{Z}$ be a realization
of the density $f_{2}$ independent of $\fh i$ for $i=1,2$. Then
for any integer $\lambda\geq0$, 
\begin{eqnarray*}
\bE\left[\gamma\left(f_{1}(\mathbf{Z}),f_{2}(\mathbf{Z})\right)\bias_{\mathbf{Z},\dist i}^{q}\left[\fh i(\mathbf{Z})\right]\right] & = & \sum_{j=1}^{\left\lfloor s\right\rfloor }c_{15,i,j,q}\left(\frac{k_{i}}{N_{i}}\right)^{\frac{j}{d}}+\sum_{m=0}^{\lambda}\sum_{\substack{j=0\\
m+j\neq0
}
}^{\lfloor s\rfloor}\frac{c_{15,i,q,j,m}}{k_{i}^{\frac{1+m}{2}}}\left(\frac{k_{i}}{N_{i}}\right)^{\frac{j}{d}}\\
 &  & +O\left(\left(\frac{k_{i}}{N_{i}}\right)^{\frac{\min(s,d)}{d}}+\frac{1}{k_{i}^{\frac{2+\lambda}{2}}}\right).
\end{eqnarray*}
\end{lem}
\begin{IEEEproof}
We use the substitution $\mathbf{T}_{i}=\Pr\left(S_{k_{i}}(\mathbf{Z})\right)$
which is the $k$th order statistic of a uniform random variable \cite{mack1979multivariate}.
Therefore, $\mathbf{T}_{i}$ has a beta distribution with parameters
$k_{i}$ and $N_{i}-k_{i}+1$. This gives 
\begin{eqnarray*}
\lefteqn{\bE\left[\gamma\left(f_{1}(\mathbf{Z}),f_{2}(\mathbf{Z})\right)\bias_{\mathbf{Z},\dist i}^{q}\left[\fh i(\mathbf{Z})\right]\right]}\\
 & = & (k_{i}-1)\binom{N_{i}-1}{k_{i}-1}\int_{\mathcal{S}}\int_{0}^{1}t^{k-1}(1-t)^{n-k}\bias_{\mathbf{Z},\dist i}^{q}\left[\fh i(\mathbf{Z})\right]dtf_{i}(Z)\gamma\left(f_{1}(Z),f_{2}(Z)\right)dZ\\
 & = & (k_{i}-1)\binom{N_{i}-1}{k_{i}-1}\int_{0}^{1}t^{k-1}(1-t)^{n-k}\int_{\mathcal{S}}\bias_{\mathbf{Z},\dist i}^{q}\left[\fh i(\mathbf{Z})\right]f_{i}(Z)\gamma\left(f_{1}(Z),f_{2}(Z)\right)dZdt\\
 & = & (k_{i}-1)\binom{N_{i}-1}{k_{i}-1}\int_{0}^{1}t^{k-1}(1-t)^{n-k}\int_{\mathcal{S}_{I_{i}}}\bias_{\mathbf{Z},\dist i}^{q}\left[\fh i(\mathbf{Z})\right]f_{i}(Z)\gamma\left(f_{1}(Z),f_{2}(Z)\right)dZdt\\
 &  & +(k_{i}-1)\binom{N_{i}-1}{k_{i}-1}\int_{0}^{1}t^{k-1}(1-t)^{n-k}\int_{\mathcal{S}_{B_{i}}}\bias_{\mathbf{Z},\dist i}^{q}\left[\fh i(\mathbf{Z})\right]f_{i}(Z)\gamma\left(f_{1}(Z),f_{2}(Z)\right)dZdt.
\end{eqnarray*}
Note that $\mathbf{T}_{i}$ monotonically increases with $\dist i$
and is therefore invertible. Thus $\dist i$ and $\mathbf{T}_{i}$
are deterministically related and $\dist i$ can be viewed as a function
of $\mathbf{T}_{i}$. Thus we can consider separately the cases where
$\mathbf{Z}$ is in $\mathcal{S}_{I_{i}}$ and $\mathcal{S}_{B_{i}}$
even after making the change of variables.

We first consider $\mathbf{Z}\in\mathcal{S}_{I_{i}}$. It is clear
in this case by (\ref{eq:conditional_knn2}) and the symmetry of $K(u)$
that 
\[
\bE_{\mathbf{Z},\dist i}\left[\fh i(\mathbf{Z})\right]=\frac{k_{i}-1}{N_{i}}\frac{1}{\Pr\left(S_{k_{i}}(\mathbf{Z})\right)}\left(f_{i}(\mathbf{Z})+\sum_{j=1}^{\left\lfloor s/2\right\rfloor }c_{i,j}(\mathbf{Z})\dist i^{2j}+O\left(\dist i^{s}\right)\right).
\]
For $q\geq2$, we obtain by the binomial theorem, 
\begin{eqnarray*}
\left(\bE_{\mathbf{Z},\dist i}\fh i(\mathbf{Z})\right)^{j} & = & \left(\frac{k_{i}-1}{N_{i}}\frac{1}{\Pr\left(S_{k_{i}}(\mathbf{Z})\right)}\right)^{j}\left(f_{i}^{j}(\mathbf{Z})+\sum_{n=1}^{\left\lfloor s/2\right\rfloor }c_{i,j,n}(\mathbf{Z})\dist i^{2n}+O\left(\dist i^{s}\right)\right),\\
\bias_{\mathbf{Z},\dist i}^{q}\left[\fh i(\mathbf{Z})\right] & = & \sum_{j=0}^{q}\binom{q}{j}\left(\bE_{\mathbf{Z},\dist i}\fh i(\mathbf{Z})\right)^{j}\left(f_{i}(\mathbf{Z})\right)^{q-j}(-1)^{j}\\
 & = & \sum_{j=0}^{q}\binom{q}{j}\left(\frac{k_{i}-1}{N_{i}}\frac{1}{\Pr\left(S_{k_{i}}(\mathbf{Z})\right)}\right)^{j}(-1)^{j}\\
 &  & \times\left(f_{i}^{q}(\mathbf{Z})+\sum_{n=1}^{\left\lfloor s/2\right\rfloor }c_{i,j,n}(\mathbf{Z})f_{i}(\mathbf{Z})^{q-j}\dist i^{2n}+O\left(\dist i^{s}\right)\right).
\end{eqnarray*}
By applying concentration inequality arguments \cite{sricharan2012thesis},
it can be shown that with high probability, 
\begin{equation}
\left(\frac{k_{i}-1}{N_{i}}\frac{1}{\Pr\left(S_{k_{i}}(\mathbf{Z})\right)}\right)^{j}=\theta\left(\frac{1}{\left(1+\frac{\sqrt{6}}{\sqrt{k_{i}}}\right)^{j}}\right).\label{eq:concentration}
\end{equation}
Then applying the binomial theorem in reverse gives (with high probability)
\begin{eqnarray*}
\sum_{j=0}^{q}\binom{q}{j}\left(\frac{k_{i}-1}{N_{i}}\frac{1}{\Pr\left(S_{k_{i}}(\mathbf{Z})\right)}\right)^{j}(-1)^{j} & = & \left(1-\frac{1}{1+\frac{\sqrt{6}}{\sqrt{k_{i}}}}\right)^{q}\\
 & = & \left(\frac{6}{k_{i}}\right)^{\frac{q}{2}}\frac{1}{\left(1+\sqrt{\frac{6}{k_{i}}}\right)^{q}}\\
 & = & \left(\frac{6}{k_{i}}\right)^{\frac{q}{2}}\sum_{j=0}^{\infty}\binom{-q}{j}(-1)^{j}\left(\frac{6}{k_{i}}\right)^{\frac{j}{2}}\\
 & = & \sum_{j=0}^{\lambda-1}\theta\left(\frac{1}{k_{i}^{\frac{q+j}{2}}}\right)+O\left(\frac{1}{k_{i}^{\frac{q+\lambda}{2}}}\right),
\end{eqnarray*}
where $\lambda$ is any nonnegative integer. Thus 
\[
\bE\left[\sum_{j=0}^{q}\binom{q}{j}\left(\frac{k_{i}-1}{N_{i}}\frac{1}{\Pr\left(S_{k_{i}}(\mathbf{Z})\right)}\right)^{j}(-1)^{j}f_{i}^{q}(\mathbf{Z})\right]=\sum_{j=0}^{\lambda-1}c_{3,i,j,q}\frac{1}{k_{i}^{\frac{q+j}{2}}}+O\left(\frac{1}{k_{i}^{\frac{q+\lambda}{2}}}\right).
\]

For $q=1$, we have 
\[
(k_{i}-1)\binom{N_{i}-1}{k_{i}-1}\int_{0}^{1}t^{k_{i}-2}(1-t)^{n-k_{i}}\int_{\mathcal{S}_{I_{i}}}f_{i}(Z)f_{2}(Z)dzdt-\int_{\mathcal{S}_{I_{i}}}f_{i}(Z)f_{2}(Z)dz=0.
\]

For the terms that include $\dist i^{\lambda}$ for some positive
integer $\lambda$, we have for $\mathbf{Z}\in\mathcal{S}_{I_{i}}$
that 
\[
\bE\left[\dist i^{\lambda}\frac{k_{i}-1}{N_{i}}\frac{1}{\Pr\left(S_{k_{i}}(\mathbf{Z})\right)}\right]=(k_{i}-1)\binom{N_{i}-1}{k_{i}-1}\int_{0}^{1}t^{k_{i}-2}(1-t)^{n-k_{i}}\int_{\mathcal{S}_{I_{i}}}\dist i^{\lambda}f_{2}(Z)dZdt.
\]
We now find an expression for $\dist i$ in terms of $\mathbf{T}_{i}$
when $\mathbf{Z}\in\mathcal{S}_{I_{i}}$. Recall that $\mathbf{T}_{i}=\Pr\left(S_{k_{i}}(\mathbf{Z})\right)$.
By Taylor series expansion, 
\begin{eqnarray}
\mathbf{T}_{i} & = & \int_{S_{k_{i}}(\mathbf{Z})}f_{i}(x)dx\nonumber \\
 & = & \dist i^{d}\left(f_{i}(\mathbf{Z})c_{d}+\sum_{j=1}^{\left\lfloor s/2\right\rfloor }c_{4,i,j}(\mathbf{Z})\dist i^{2j}+O\left(\dist i^{s}\right)\right)\nonumber \\
\implies\dist i & = & \frac{\mathbf{T}_{i}^{\frac{1}{d}}}{\left(f_{i}(\mathbf{Z})c_{d}+\sum_{j=1}^{\left\lfloor s/2\right\rfloor }c_{4,i,j}(\mathbf{Z})\dist i^{2j}+O\left(\dist i^{s}\right)\right)^{\frac{1}{d}}}.\label{eq:dist2t}
\end{eqnarray}
Note that as $\dist i\downarrow0$, we have that $\left|\sum_{j=1}^{\left\lfloor s/2\right\rfloor }c_{4,i,j}(\mathbf{Z})\dist i^{2j}+O\left(\dist i^{s}\right)\right|<f_{i}(\mathbf{Z})c_{d}$
for sufficiently small $\dist i$ since we assume that $f_{i}(x)\geq\epsilon_{0}>0$.
Therefore, we can apply the generalized binomial theorem to obtain
\begin{eqnarray*}
\left(f_{i}(\mathbf{Z})c_{d}+\sum_{j=1}^{\left\lfloor s/2\right\rfloor }c_{4,i,j}(\mathbf{Z})\dist i^{2j}+O\left(\dist i^{s}\right)\right)^{-\frac{1}{d}} & = & \sum_{m=0}^{\infty}\binom{-1/d}{m}\left(f_{i}(\mathbf{Z})c_{d}\right)^{-1/d-m}\\
 &  & \times\left(\sum_{j=1}^{\left\lfloor s/2\right\rfloor }c_{4,i,j}(\mathbf{Z})\dist i^{2j}+O\left(\dist i^{s}\right)\right)^{m}\\
 & = & \left(f_{i}(\mathbf{Z})c_{d}\right)^{-1/d}+\sum_{j=1}^{\left\lfloor s/2\right\rfloor }c_{5,i,j}(\mathbf{Z})\dist i^{2j}\\
 &  & +O\left(\dist i^{s}\right).
\end{eqnarray*}
Using this expression in (\ref{eq:dist2t}) and resubstituting the
LHS into the RHS gives that 
\begin{eqnarray*}
\dist i & = & \left(\frac{\mathbf{T}_{i}}{f_{i}(\mathbf{Z})c_{d}}\right)^{\frac{1}{d}}+\sum_{j=1}^{\left\lfloor s/2\right\rfloor }c_{6,i,j}(\mathbf{Z})\mathbf{T}_{i}^{2j/d}+O\left(\mathbf{T}_{i}^{s/d}\right),\\
\implies\dist i^{\lambda} & = & \left(\frac{\mathbf{T}_{i}}{f_{i}(\mathbf{Z})c_{d}}\right)^{\frac{\lambda}{d}}+\sum_{j=1}^{\left\lfloor s/2\right\rfloor }c_{7,i,j}(\mathbf{Z})\mathbf{T}_{i}^{2j\lambda/d}+O\left(\mathbf{T}_{i}^{s\lambda/d}\right).
\end{eqnarray*}
Therefore, 
\begin{eqnarray*}
\lefteqn{\bE\left[\dist i^{\lambda}\frac{k_{i}-1}{N_{i}}\frac{1}{\Pr\left(S_{k_{i}}(\mathbf{Z})\right)}\right]}\\
 & = & (k_{i}-1)\binom{N_{i}-1}{k_{i}-1}\int_{0}^{1}t^{k_{i}-2+\lambda/d}(1-t)^{n-k_{i}}\int_{\mathcal{S}_{I_{i}}}\frac{f_{2}(Z)}{\left(f_{i}(Z)c_{d}\right)^{\lambda/d}}dZdt\\
 &  & +\sum_{j=1}^{\left\lfloor s/2\right\rfloor }(k_{i}-1)\binom{N_{i}-1}{k_{i}-1}\int_{0}^{1}t^{k_{i}-2+2j\lambda/d}(1-t)^{n-k_{i}}\int_{\mathcal{S}_{I_{i}}}f_{2}(Z)c_{7,i,j}(Z)dZdt\\
 & = & c_{7,i,0}\left(\frac{k_{i}}{N_{i}}\right)^{\lambda/d}+\sum_{j=1}^{\left\lfloor s/2\right\rfloor }c_{7,i,j}\left(\frac{k_{i}}{N_{i}}\right)^{2\lambda j/d}+O\left(\left(\frac{k_{i}}{N_{i}}\right)^{\frac{s}{d}}\right).
\end{eqnarray*}
Combining this result with (\ref{eq:concentration}) gives for $q\geq2$
and any integer $\lambda\geq0$ 
\[
N_{i}\binom{N_{i}-1}{k_{i}-1}\int_{0}^{1}t^{k_{i}-2}(1-t)^{N_{i}-k_{i}}\int_{\mathcal{S}_{I_{i}}}\bias_{\mathbf{Z},\dist i}^{q}\left[\fh i(\mathbf{Z})\right]f_{i}(Z)\gamma\left(f_{1}(Z),f_{2}(Z)\right)dZdt
\]
\[
=\sum_{j=0}^{\lambda-1}c_{3,i,j,q}\frac{1}{k^{\frac{q+j}{2}}}+O\left(\frac{1}{k^{\frac{q+\lambda}{2}}}+\left(\frac{k_{i}}{N_{i}}\right)^{\frac{s}{d}}\right)+\sum_{m=0}^{\lambda-1}\sum_{j=1}^{\left\lfloor s/2\right\rfloor }c_{7,i,j,m,q}\left(\frac{k_{i}}{N_{i}}\right)^{2j/d}\frac{1}{k_{i}^{\frac{q-1+m}{2}}}.
\]
Similarly, for $q=1,$ 
\[
N_{i}\binom{N_{i}-1}{k_{i}-1}\int_{0}^{1}t^{k_{i}-2}(1-t)^{N_{i}-k_{i}}\int_{\mathcal{S}_{I_{i}}}\bias_{\mathbf{Z},\dist i}\left[\fh i(\mathbf{Z})\right]f_{i}(Z)\gamma\left(f_{1}(Z),f_{2}(Z)\right)dZdt
\]
\[
=\sum_{j=1}^{\left\lfloor s/2\right\rfloor }c_{7,i,j,m,1}\left(\frac{k_{i}}{N_{i}}\right)^{2j/d}+O\left(\left(\frac{k_{i}}{N_{i}}\right)^{\frac{s}{d}}\right).
\]

We now consider the case where $\mathbf{Z}\in\mathcal{S}_{B_{i}}$.
In this case, we extend the density beyond the boundary. This gives
\begin{eqnarray*}
\bias_{\mathbf{Z},\dist i}\left[\fh i(\mathbf{Z})\right] & = & \frac{k_{i}-1}{N_{i}}\frac{1}{\dist i^{d}}\frac{1}{\Pr\left(S_{k_{i}}(\mathbf{Z})\right)}\int_{S_{k_{i}}(\mathbf{Z})\cap\mathcal{S}}K\left(\frac{\mathbf{Z}-x}{\dist i}\right)f_{i}(x)dx\\
 & = & \frac{k_{i}-1}{N_{i}}\frac{1}{\dist i^{d}}\frac{1}{\Pr\left(S_{k_{i}}(\mathbf{Z})\right)}\int_{S_{k_{i}}(\mathbf{Z})}K\left(\frac{\mathbf{Z}-x}{\dist i}\right)f_{i}(x)dx-f_{i}(\mathbf{Z})\\
 &  & -\frac{k_{i}-1}{N_{i}}\frac{1}{\dist i^{d}}\frac{1}{\Pr\left(S_{k_{i}}(\mathbf{Z})\right)}\int_{x\notin\mathcal{S}}K\left(\frac{\mathbf{Z}-x}{\dist i}\right)f_{i}(x)dx\\
 & = & T_{1}(\mathbf{Z},\dist i)-T_{2}(\mathbf{Z},\dist i).
\end{eqnarray*}
The expression for $T_{1}(\mathbf{Z},\dist i)$ is identical to that
when $\mathbf{Z}\in\mathcal{S}_{I_{i}}$ and so taking the expectation
gives the same results. Therefore, we focus on $T_{2}(\mathbf{Z},\dist i)$.
As before, we substitute $u=(\mathbf{Z}-x)/\dist i$ inside the integral
and take a Taylor series expansion of $f_{i}$ to get 
\[
\sum_{|\alpha|\leq\left\lfloor s\right\rfloor }\frac{D^{\alpha}f_{i}(\mathbf{Z})}{\alpha!}\dist i^{d+|\alpha|}\int_{u:\mathbf{Z}+u\dist i\notin\mathcal{S}}u^{\alpha}K(u)du+O\left(\dist i^{d+s}\right).
\]
As before, we can again substitute $\mathbf{T}_{i}=\Pr\left(S_{k_{i}}(\mathbf{Z})\right)$.
However, we need to find an expression for $\dist i$ in terms of
$\mathbf{T}_{i}$ for $\mathbf{Z}\in\mathcal{S}_{B_{i}}$. Note that
\begin{eqnarray}
\mathbf{T}_{i} & = & \int_{z\in S_{k_{i}}(\mathbf{Z})\cap\mathcal{S}}f_{i}(z)dz.\nonumber \\
 & = & \int_{z\in S_{k_{i}}(\mathbf{Z})}f(z)dz-\int_{z\in S_{k_{i}}(\mathbf{Z})\cap\mathcal{S}^{C}}f(z)dz\nonumber \\
 & = & \dist i^{d}\left(f_{i}(\mathbf{Z})c_{d}+\sum_{j=1}^{\left\lfloor s/2\right\rfloor }c_{4,i,j}(\mathbf{Z})\dist i{}^{2j}+O(\dist i^{s})\right)\nonumber \\
 &  & -\int_{z\in S_{k_{i}}(\mathbf{Z})\cap\mathcal{S}^{C}}\left(\sum_{|\alpha|\leq\left\lfloor s\right\rfloor }\frac{(z-\mathbf{Z})^{\alpha}}{\alpha!}D^{\alpha}f(\mathbf{Z})+O\left(\left(z-\mathbf{Z}\right)^{s}\right)\right)dz.\label{eq:t2r}
\end{eqnarray}
We need to simplify the second integral in (\ref{eq:t2r}) before
solving for $\dist i$. If we assume that the support $\mathcal{S}=[0,1]^{d}$,
then we can use the techniques used in Appendix~\ref{sec:boundary_sphere}.

Assume that $d$ is odd as as the case for even $d$ will be similar.
We first consider the case where only a single coordinate $\mathbf{Z}_{(1)}$
is close to the boundary. Without loss of generality, we assume that
$\mathbf{Z}_{(1)}$ is close to 1. Then for a given $\alpha$, we
can use (\ref{eq:polyodd}) to obtain 
\begin{align}
\int_{z\in S_{k_{i}}(\mathbf{Z})\cap\mathcal{S}^{C}}\frac{(z-\mathbf{Z})^{\alpha}}{\alpha!}D^{\alpha}f_{i}(\mathbf{Z})dz=\dist i^{d+|\alpha|}\sum_{m=0}^{|\alpha|}\sum_{\ell=0}^{d+|\alpha|}p_{m,\ell,\alpha,i}(\mathbf{Z})\left(\sqrt{1-\left(\frac{1-\mathbf{Z}_{(1)}}{\dist i}\right)^{2}}\right)^{\ell}\nonumber \\
\times\left(\frac{1-\mathbf{Z}_{(1)}}{\dist i}\right)^{m},\label{eq:tboundary_int}
\end{align}
where $p_{m,\ell,\alpha,i}(\mathbf{Z})$ is $\left\lfloor s\right\rfloor -|\alpha|$
times differentiable wrt $\mathbf{Z}$. Now expand $p_{m,\ell,\alpha,i}(\mathbf{Z})$
only in the $\mathbf{Z}_{(1)}$ coordinate at $\mathbf{Z}_{(1)}=1$
to get 
\[
p_{m,\ell,\alpha,i}(\mathbf{Z})=\sum_{j=0}^{\left\lfloor s\right\rfloor -|\alpha|}\frac{\partial^{j}p_{m,\ell,\alpha,i}(1,\mathbf{Z}_{(2)},\dots,\mathbf{Z}_{(d)})}{\partial\mathbf{Z}_{(1)}^{j}}\frac{\left(1-\mathbf{Z}_{(1)}\right)^{n}}{j!}.
\]
Substituting this into (\ref{eq:tboundary_int}) and substituting
$\mathbf{W}=\frac{1-\mathbf{Z}_{(1)}}{\dist i}$ gives 
\[
\sum_{m=0}^{|\alpha|}\sum_{\ell=0}^{d+|\alpha|}\sum_{j=0}^{\left\lfloor s\right\rfloor -|\alpha|}\frac{\partial^{j}p_{m,\ell,\alpha,i}(1,\mathbf{Z}_{(2)},\dots,\mathbf{Z}_{(d)})}{\partial\mathbf{Z}_{(1)}^{j}}\frac{1}{j!}\left(\sqrt{1-\left(\frac{1-\mathbf{Z}_{(1)}}{\dist i}\right)^{2}}\right)^{\ell}\left(\frac{1-\mathbf{Z}_{(1)}}{\dist i}\right)^{m+j}\dist i^{j+d+|\alpha|}
\]
\[
=\sum_{m=0}^{|\alpha|}\sum_{\ell=0}^{d+|\alpha|}\sum_{j=0}^{\left\lfloor s\right\rfloor -|\alpha|}p_{m,\ell,\alpha,i}^{'}(\mathbf{Z}^{'})\left(\sqrt{1-\mathbf{W}^{2}}\right)^{\ell}\mathbf{W}^{m+j}\dist i^{j+d+|\alpha|},
\]
where $\mathbf{Z}^{'}=(1,\mathbf{Z}_{(2)},\dots,\mathbf{Z}_{(d)})$
and $p_{m,\ell,\alpha,i}^{'}(\mathbf{Z}^{'})=\frac{\partial^{j}p_{m,\ell,\alpha,i}(1,\mathbf{Z}_{(2)},\dots,\mathbf{Z}_{(d)})}{\partial\mathbf{Z}_{(1)}^{j}}\frac{1}{j!}$.
The variable $\mathbf{W}$ ranges from $0$ to $1$. Thus we have
separated the dependence on $\dist i$. Substituting these results
into (\ref{eq:t2r}) gives 
\begin{eqnarray*}
\mathbf{T}_{i} & = & \dist i^{d}\left(f_{i}(\mathbf{Z})c_{d}+\sum_{j=1}^{\left\lfloor s/2\right\rfloor }c_{4,i,j}(\mathbf{Z})\dist i{}^{2j}+O(\dist i^{s})\right)\\
 &  & -\dist i^{d}\sum_{|\alpha|\leq\left\lfloor s\right\rfloor }\sum_{m=0}^{|\alpha|}\sum_{\ell=0}^{d+|\alpha|}\sum_{j=0}^{\left\lfloor s\right\rfloor -|\alpha|}p_{m,\ell,\alpha,i}^{'}(\mathbf{Z}^{'})\left(\sqrt{1-\mathbf{W}^{2}}\right)^{\ell}\mathbf{W}^{m+j}\dist i^{j+|\alpha|}.
\end{eqnarray*}
By substituting $\mathbf{Z}_{(1)}=1-\mathbf{W}\dist i$ in the first
term and taking a Taylor series expansion of $f_{i}(\mathbf{Z})^{-1/d}$
and $c_{4,i,j}(\mathbf{Z})$ at $\mathbf{Z}_{(1)}=1$ gives 
\[
\sum_{j=0}^{\left\lfloor s\right\rfloor }c_{8,i,j}(\mathbf{Z}^{''})\dist i^{j+d}+O\left(\dist i^{s+d}\right),
\]
where $\mathbf{Z}^{''}=(\mathbf{W},\mathbf{Z}_{(2)},\dots,\mathbf{Z}_{(d)})$.
Thus we can write 
\begin{eqnarray}
\mathbf{T}_{i} & = & \dist i^{d}\left(\sum_{j=0}^{\left\lfloor s\right\rfloor }c_{9,i,j}(\mathbf{Z}^{''})\dist i^{j}+O\left(\dist i^{s}\right)\right)\nonumber \\
\implies\dist i & = & \frac{t^{\frac{1}{d}}}{\left(\sum_{j=0}^{\left\lfloor s\right\rfloor }c_{9,i,j}(\mathbf{Z}^{''})\dist i^{j}+O\left(\dist i^{s}\right)\right)^{\frac{1}{d}}}.\label{eq:r2t}
\end{eqnarray}
Then since $\dist i\downarrow0$, applying the generalized binomial
theorem to the denominator gives 
\begin{eqnarray*}
\lefteqn{\left(\sum_{j=0}^{\left\lfloor s\right\rfloor }c_{9,i,j}(\mathbf{Z}^{''})\dist i^{j}+O\left(\dist i^{s}\right)\right)^{-\frac{1}{d}}}\\
 & = & \sum_{m=0}^{\infty}\binom{-1/d}{m}c_{9,i,0}(\mathbf{Z}^{''})^{-\frac{1}{d}-j}\left(\sum_{j=1}^{\left\lfloor s\right\rfloor }c_{9,i,j}(\mathbf{Z}^{''})\dist i^{j}+O\left(\dist i^{s}\right)\right)^{m}\\
 & = & c_{9,i,0}(\mathbf{Z}^{''})^{-\frac{1}{d}}+\sum_{j=1}^{\left\lfloor s\right\rfloor }c_{10,i,j}(\mathbf{Z}^{''})\dist i^{j}+O\left(\dist i^{s}\right).
\end{eqnarray*}
Applying this result to (\ref{eq:r2t}) gives 
\begin{equation}
\dist i=\left(\frac{\mathbf{T}_{i}}{c_{9,i,0}(\mathbf{Z}^{''})}\right)^{\frac{1}{d}}+\mathbf{T}_{i}^{\frac{1}{d}}\sum_{j=1}^{\left\lfloor s\right\rfloor }c_{10,i,j}(\mathbf{Z}^{''})\dist i^{j}+O\left(\mathbf{T}_{i}^{\frac{1}{d}}\dist i^{s}\right).\label{eq:r2t2}
\end{equation}
Resubstituting the LHS of (\ref{eq:r2t2}) into the RHS multiple times
then gives 
\begin{eqnarray*}
\dist i & = & \sum_{j=1}^{\left\lfloor s\right\rfloor }c_{11,i,j}(\mathbf{Z}^{''})\mathbf{T}_{i}^{\frac{j}{d}}+O\left(\mathbf{T}_{i}^{\frac{s}{d}}\right)\\
\implies\dist i^{\lambda} & = & \sum_{j=1}^{\left\lfloor s\right\rfloor }c_{12,i,j,\lambda}(\mathbf{Z}^{''})\mathbf{T}_{i}^{\frac{j\lambda}{d}}+O\left(\mathbf{T}_{i}^{\frac{s\lambda}{d}}\right).
\end{eqnarray*}
Given these results and the fact that $\mathbf{T}_{i}$ has a beta
distribution, we have that 
\begin{align*}
\lefteqn{\bE\left[1_{\{\mathbf{Z}\in\mathcal{S}_{B_{i}}\}}\bE_{\mathbf{Z},\dist i}\left[\frac{k_{i}-1}{N_{i}}\frac{1}{\Pr\left(S_{k_{i}}(\mathbf{Z})\right)}\dist i^{\lambda}\right]\right]}\\
 & = & \frac{k_{i}-1}{N_{i}}\binom{N_{i}-1}{k_{i}-1}\int_{0}^{1}t^{k-2}(1-t)^{n-k}\int_{\mathcal{S}}\dist i^{\lambda}f_{i}(Z)dZdt.
\end{align*}
Taking a Taylor series expansion of $f_{i}$ at $Z_{(1)}=1$ gives
\[
f_{i}(Z)=\sum_{j=0}^{\left\lfloor s\right\rfloor }\frac{\partial^{j}f_{i}(Z^{'})}{\partial Z_{(1)}^{j}}W^{j}\dist i^{j}+O\left(\dist i^{s}\right).
\]
Combining all of these results gives that $\bE\left[1_{\{\mathbf{Z}\in\mathcal{S}_{B_{i}}\}}\bE_{\mathbf{Z},\dist i}\left[\frac{k_{i}-1}{N_{i}}\frac{1}{\Pr\left(S_{k_{i}}(\mathbf{Z})\right)}\dist i^{\lambda}\right]\right]$
has terms of the form of 
\[
(k_{i}-1)\binom{N_{i}-1}{k_{i}-1}\int_{0}^{1}t^{k-2+\frac{\lambda+1}{d}}(1-t)^{n-k}dt=\left(\frac{k_{i}}{N_{i}}\right)^{\frac{\lambda+1}{d}}+o\left(\frac{k_{i}}{N_{i}}\right).
\]
Therefore, 
\begin{eqnarray}
\bE\left[T_{2}(\mathbf{Z},\dist i)\right] & = & \left(k_{i}-1\right)\binom{N_{i}-1}{k_{i}-1}\int_{0}^{1}t^{k_{i}-2}(1-t)^{N_{i}-k_{i}}\nonumber \\
 &  & \times\int_{S_{B_{i}}}\left(\sum_{j=0}^{\left\lfloor s\right\rfloor }c_{13,i,j}(Z^{''})t^{\frac{j+1}{d}}+O\left(t^{\frac{s}{d}}\right)\right)dZ^{''}dt\nonumber \\
 & = & \sum_{j=1}^{\left\lfloor s\right\rfloor }c_{14,i,j}\left(\frac{k_{i}}{N_{i}}\right)^{\frac{j}{d}}+O\left(\left(\frac{k_{i}}{N_{i}}\right)^{\min(s,d)/d}\right).\label{eq:boundary_final}
\end{eqnarray}

For $\bE\left[\left(T_{1}(\mathbf{Z},\dist i)-T_{2}(\mathbf{Z},\dist i)\right)^{q}\right]$,
we have by the binomial theorem that 
\[
\left(T_{1}(\mathbf{Z},\dist i)-T_{2}(\mathbf{Z},\dist i)\right)^{q}=\sum_{j=0}^{q}\binom{q}{j}T_{1}(\mathbf{Z},\dist i)^{j}T_{2}(\mathbf{Z},\dist i)^{q-j}.
\]
Applying a similar analysis gives similar results.

For the case when $S_{k_{i}}(\mathbf{Z})$ intersects multiple boundary
points, a similar approach can be used as in Appendix~\ref{sec:bound_sphere_mult}.
This will yield a similar expression to (\ref{eq:boundary_final}).
Combining all results with the fact that $\gamma(x,y)$ is bounded
finishes the proof.
\end{IEEEproof}
\begin{lem}
\label{lem:exp_bias2}Let $\gamma(x,y)$ be an arbitrary function
satisfying $\sup_{x,y}|\gamma(x,y)|<\infty$. Let $\mathbf{Z}$ be
a realization of the density $f_{2}$ independent of $\fh i$ for
$i=1,2$. Then for any integer $\lambda\geq0$ 
\begin{eqnarray*}
\lefteqn{\bE\left[\gamma(f_{1}(\mathbf{Z}),f_{2}(\mathbf{Z}))\bias_{\mathbf{Z},\dist 1}^{t}\left[\fh 1(\mathbf{Z})\right]\bias_{\mathbf{Z},\dist 2}^{q}\left[\fh 2(\mathbf{Z})\right]\right]}\\
 & = & \sum_{j=0}^{\lfloor s\rfloor}\sum_{\substack{i=0\\
i+j\neq0
}
}^{\lfloor s\rfloor}c_{16,i,j,q,t}\left(\frac{k_{1}}{N_{1}}\right)^{\frac{i}{d}}\left(\frac{k_{2}}{N_{2}}\right)^{\frac{j}{d}}+O\left(\max\left(\frac{k_{1}}{N_{1}},\frac{k_{2}}{N_{2}}\right)^{\frac{\min(s,d)}{d}}+\frac{1}{\min(k_{1},k_{2})^{\frac{2+\lambda}{2}}}\right)\\
 &  & +\sum_{m=0}^{\lambda}\sum_{\substack{j=0\\
m+j\neq0
}
}^{\lfloor s\rfloor}\sum_{n=0}^{\lambda}\sum_{\substack{i=0\\
n+i\neq0
}
}^{\lfloor s\rfloor}\frac{c_{16,i,j,q,t,m,n}}{k_{1}^{\frac{1+m}{2}}k_{2}^{\frac{1+n}{2}}}\left(\frac{k_{1}}{N_{1}}\right)^{\frac{i}{d}}\left(\frac{k_{2}}{N_{2}}\right)^{\frac{j}{d}}.
\end{eqnarray*}
\end{lem}
\begin{IEEEproof}
Note that $\dist 1$ and $\dist 2$ are conditionally independent
of each other given $\mathbf{Z}$. Applying similar techniques as
in the proof of Lemma~\ref{lem:exp_bias} yields the result.
\end{IEEEproof}
Applying Lemmas~\ref{lem:exp_bias} and \ref{lem:exp_bias2} to (\ref{eq:bias_term})
gives 
\begin{eqnarray}
\lefteqn{\bE\left[g\left(\bE_{\mathbf{Z},\dist 1(\mathbf{Z})}\fh 1(\mathbf{Z}),\bE_{\mathbf{Z},\dist 2(\mathbf{Z})}\fh 2(\mathbf{Z})\right)-g\left(f_{1}(\mathbf{Z}),f_{2}(\mathbf{Z})\right)\right]}\nonumber \\
 & = & \sum_{j=0}^{\lfloor s\rfloor}\sum_{\substack{i=0\\
i+j\neq0
}
}^{\lfloor s\rfloor}c_{18,i,j}\left(\frac{k_{1}}{N_{1}}\right)^{\frac{i}{d}}\left(\frac{k_{2}}{N_{2}}\right)^{\frac{j}{d}}+O\left(\max\left(\frac{k_{1}}{N_{1}},\frac{k_{2}}{N_{2}}\right)^{\frac{\min(s,d)}{d}}+\frac{1}{\min(k_{1},k_{2})^{\frac{2+\lambda}{2}}}\right)\nonumber \\
 &  & +\sum_{m=0}^{\lambda}\sum_{\substack{j=0\\
j+m\neq0
}
}^{r}\left(\frac{c_{17,1,j,m}}{k_{1}^{\frac{1+m}{2}}}\left(\frac{k_{1}}{N_{1}}\right)^{\frac{j}{d}}+\frac{c_{17,2,j,m}}{k_{2}^{\frac{1+m}{2}}}\left(\frac{k_{2}}{N_{2}}\right)^{\frac{j}{d}}\right)\nonumber \\
 &  & +\sum_{j=1}^{r}\left(c_{17,1,j}\left(\frac{k_{1}}{N_{1}}\right)^{\frac{j}{d}}+c_{17,2,j}\left(\frac{k_{2}}{N_{2}}\right)^{\frac{j}{d}}\right)\nonumber \\
 &  & +\sum_{m=0}^{\lambda}\sum_{\substack{j=0\\
m+j\neq0
}
}^{\lfloor s\rfloor}\sum_{n=0}^{\lambda}\sum_{\substack{i=0\\
n+i\neq0
}
}^{\lfloor s\rfloor}\frac{c_{18,i,j,m,n}}{k_{1}^{\frac{1+m}{2}}k_{2}^{\frac{1+n}{2}}}\left(\frac{k_{1}}{N_{1}}\right)^{\frac{i}{d}}\left(\frac{k_{2}}{N_{2}}\right)^{\frac{j}{d}}.\label{eq:bias_terms}
\end{eqnarray}

We now focus on the first term in (\ref{eq:gsplit}). The truncated
Taylor series expansion of $g\left(\fh 1(\mathbf{Z}),\fh 2(\mathbf{Z})\right)$
around $\bE_{\mathbf{Z},\dist 1}\fh 1(\mathbf{Z})$ and $\bE_{\mathbf{Z},\dist 2}\fh 2(\mathbf{Z})$
gives 
\begin{align}
\lefteqn{g\left(\fh 1(\mathbf{Z}),\fh 2(\mathbf{Z})\right)}\nonumber \\
 & = & \sum_{i=0}^{\nu}\sum_{j=0}^{\nu}\left(\left.\frac{\partial^{i+j}g(x,y)}{\partial x^{i}\partial y^{j}}\right|_{\substack{x=\bE_{\mathbf{Z},\dist 1}\fh 1(\mathbf{Z})\\
y=\bE_{\mathbf{Z},\dist 2}\fh 2(\mathbf{Z})
}
}\right)\frac{\ehat 1^{i}(\mathbf{Z})\ehat 2^{j}(\mathbf{Z})}{i!j!}+o\left(\ehat 1^{\nu}(\mathbf{Z})+\ehat 2^{\nu}(\mathbf{Z})\right),\label{eq:gvariance}
\end{align}
where $\ehat i:=\fh i(\mathbf{Z})-\bE_{\mathbf{Z},\dist i}\fh i(\mathbf{Z})$.
We thus require expressions for $\bE_{\mathbf{Z},\dist i}\left[\ehat i^{j}(\mathbf{Z})\right]$
to control this expression.
\begin{lem}
\label{lem:var_term}Let $\mathbf{Z}$ be a realization of the density
$f_{2}$ that is in the interior of the support wrt $\dist i$ and
is independent of $\fh i$ for $i=1,2$. Let $n(q)$ be the set of
integer divisors of $q$ including 1 but excluding q. Then, 
\[
\bE_{\mathbf{Z},\dist i}\left[\ehat i^{q}(\mathbf{Z})\right]=\begin{cases}
\frac{k_{i}-1}{N_{i}\Pr\left(S_{k_{i}}(\mathbf{Z})\right)}\sum_{j\in n(q)}\frac{1}{\left(N_{i}\dist i^{d}\right)^{q-j}}\sum_{m=0}^{\lfloor s/2\rfloor}c_{,i,q,j,m}(\mathbf{Z})\dist i^{2m}+O\left(\frac{\dist i^{2d}}{k_{i}}\right), & q\geq2\\
0, & q=1
\end{cases}
\]
\begin{align*}
\lefteqn{\bE_{\mathbf{Z},\dist 1,\dist 2}\left[\ehat 1^{q}(\mathbf{Z})\ehat 2^{l}(\mathbf{Z})\right]}\\
 & = & \begin{cases}
\frac{k_{i}-1}{N_{i}\Pr\left(S_{k_{i}}(\mathbf{Z})\right)}\left(\sum_{j\in n(q)}\frac{1}{\left(N_{1}\dist 1^{d}\right)^{q-j}}\sum_{m=0}^{\lfloor s/2\rfloor}c_{,1,q,j,m}(\mathbf{Z})\dist 1^{2m}\right)\times & q,l\geq2\\
\left(\sum_{i\in n(l)}\frac{1}{\left(N_{2}\dist 2^{d}\right)^{l-i}}\sum_{t=0}^{\lfloor s/2\rfloor}c_{,2,l,i,t}(\mathbf{Z})\dist 2^{2t}\right)+O\left(\frac{1}{N_{1}}+\frac{1}{N_{2}}\right),\\
0, & q=1\,or\,l=1.
\end{cases}
\end{align*}
\end{lem}
\begin{IEEEproof}
Define the random variable $\mathbf{V}_{i}(\mathbf{Z})=K\left(\frac{\mathbf{X}_{i}-\mathbf{Z}}{\dist 2}\right)-\bE_{\mathbf{Z},\dist 2}K\left(\frac{\mathbf{X}_{i}-\mathbf{Z}}{\dist 2}\right)$.
Then 
\begin{eqnarray*}
\ehat 2(\mathbf{Z}) & = & \fh 2(\mathbf{Z})-\bE_{\mathbf{Z},\dist 2}\fh 2(\mathbf{Z})\\
 & = & \frac{1}{N_{2}\dist 2^{d}}\sum_{i=1}^{N_{2}}\mathbf{V}_{i}(\mathbf{Z}).
\end{eqnarray*}
Note that $\bE_{\mathbf{Z},\dist 2}\mathbf{V}_{i}(\mathbf{Z})=0$.
From our previous results, we have for $j\geq1$, 
\begin{eqnarray*}
\bE_{\mathbf{Z},\dist 2}\left[K^{j}\left(\frac{\mathbf{X}_{i}-\mathbf{Z}}{\dist 2}\right)\right] & = & \bE_{\mathbf{Z},\dist 2}\left[K\left(\frac{\mathbf{X}_{i}-\mathbf{Z}}{\dist 2}\right)\right]\\
 & = & \frac{k_{2}-1}{N_{2}}\frac{\dist 2^{d}}{\Pr\left(S_{k_{2}}(\mathbf{Z})\right)}\sum_{m=0}^{\lfloor s/2\rfloor}c_{2,m}(\mathbf{Z})\dist 2^{2m}+O\left(\dist 2^{s}\right).
\end{eqnarray*}
By the binomial theorem, 
\begin{eqnarray*}
\lefteqn{\bE_{\mathbf{Z},\dist 2}\left[\mathbf{V}_{i}^{j}(\mathbf{Z})\right]}\\
 & = & \sum_{n=0}^{j}\binom{j}{n}\bE_{\mathbf{Z},\dist 2}\left[K^{j}\left(\frac{\mathbf{X}_{i}-\mathbf{Z}}{\dist 2}\right)\right]\bE_{\mathbf{Z},\dist 2}\left[K\left(\frac{\mathbf{X}_{i}-\mathbf{Z}}{\dist 2}\right)\right]^{j-n}\\
 & = & \sum_{n=0}^{j}\binom{j}{n}\left(\frac{k_{2}-1}{N_{2}}\frac{\dist 2^{d}}{\Pr\left(S_{k_{2}}(\mathbf{Z})\right)}\sum_{m=0}^{\lfloor s/2\rfloor}c_{2,m}(\mathbf{Z})\dist 2^{2m}\right)O\left(\left(\frac{\dist 2^{d}}{\Pr\left(S_{k_{2}}(\mathbf{Z})\right)}\frac{k_{2}-1}{N_{2}}\right)^{j-n}\right)\\
 & = & \frac{k_{2}-1}{N_{2}}\frac{\dist 2^{d}}{\Pr\left(S_{k_{2}}(\mathbf{Z})\right)}\sum_{m=0}^{\lfloor s/2\rfloor}c_{2,m}(\mathbf{Z})\dist 2^{2m}+O\left(\left(\frac{\dist 2^{d}}{\Pr\left(S_{k_{2}}(\mathbf{Z})\right)}\frac{k_{2}-1}{N_{2}}\right)^{2}\right).
\end{eqnarray*}
We can use these expressions to simplify $\bE_{\mathbf{Z},\dist i}\left[\ehat i^{q}(\mathbf{Z})\right]$.
For example, let $q=2$. Due to the independence of the $\mathbf{X}_{i}$s
and the fact that with high probability 
\[
\left(\frac{1}{\Pr\left(S_{k_{i}}(\mathbf{Z})\right)}\frac{k_{i}-1}{N_{i}}\right)^{2}=O\left(\frac{1}{k_{i}}\right),
\]
we obtain 
\begin{align*}
\bE_{\mathbf{Z},\dist 2}\left[\ehat 2^{2}(\mathbf{Z})\right] & =\frac{1}{N_{2}\dist 2^{2d}}\bE_{\mathbf{Z},\dist 2}\left[\mathbf{V}_{i}^{2}(\mathbf{Z})\right]\\
 & =\frac{k_{2}-1}{N_{2}\Pr\left(S_{k_{2}}(\mathbf{Z})\right)}\frac{1}{N_{2}\dist 2^{d}}\sum_{m=0}^{\lfloor s/2\rfloor}c_{2,m}(\mathbf{Z})\dist 2^{2m}+O\left(\frac{\dist 2^{2d}}{k_{2}}\right).
\end{align*}

Similarly, for $q=3$, 
\begin{align*}
\bE_{\mathbf{Z},\dist 2}\left[\ehat 2^{3}(\mathbf{Z})\right] & =\frac{1}{N_{2}^{2}\dist 2^{3d}}\bE_{\mathbf{Z},\dist 2}\left[\mathbf{V}_{i}^{3}(\mathbf{Z})\right]\\
 & =\frac{k_{2}-1}{N_{2}\Pr\left(S_{k_{2}}(\mathbf{Z})\right)}\frac{1}{\left(N_{2}\dist 2^{d}\right)^{2}}\sum_{m=0}^{\lfloor s/2\rfloor}c_{2,m}(\mathbf{Z})\dist 2^{2m}+O\left(\frac{\dist 2^{2d}}{k_{2}}\right),
\end{align*}
and for $q=4$, 
\begin{align*}
\bE_{\mathbf{Z},\dist 2}\left[\ehat 2^{4}(\mathbf{Z})\right] & =\frac{1}{N_{2}^{3}\dist 2^{4d}}\bE_{\mathbf{Z},\dist 2}\left[\mathbf{V}_{i}^{4}(\mathbf{Z})\right]+\frac{N_{2}-1}{N_{2}^{3}\dist 2^{4d}}\left(\bE_{\mathbf{Z},\dist 2}\left[\mathbf{V}_{i}^{2}(\mathbf{Z})\right]\right)^{2}\\
 & =\frac{k_{2}-1}{N_{2}\Pr\left(S_{k_{2}}(\mathbf{Z})\right)}\left(\frac{1}{\left(N_{2}\dist 2^{d}\right)^{3}}+\frac{1}{\left(N_{2}\dist 2^{d}\right)^{2}}\right)\sum_{m=0}^{\lfloor s/2\rfloor}c_{2,m}(\mathbf{Z})\dist 2^{2m}+O\left(\frac{\dist 2^{2d}}{k_{2}}\right).
\end{align*}
It can then be seen that for $q\geq2,$ the pattern is given in the
first expression in the lemma statement.

For any integer $q$, the largest possible factor is $q/2$. Therefore,
the smallest possible exponent on the $N_{2}\dist 2^{d}$ term is
$q/2$. This increases as q increases. A similar expression for $\bE_{\mathbf{Z},\dist i}\left[\ehat i^{q}(\mathbf{Z})\right]$
for $i=1$ can be proved using a similar technique. The second expression
in the lemma statement then follows from the fact that $\ehat 1(\mathbf{Z})$
and $\ehat 2(\mathbf{Z})$ are conditionally independent given $\mathbf{Z}$,
$\dist 1$, and $\dist 2$
\end{IEEEproof}
For general $g$, we can only say that 
\[
\left.\frac{\partial^{i+j}g(x,y)}{\partial x^{i}\partial y^{j}}\right|_{\substack{x=\bE_{\mathbf{Z},\dist 1}\fh 1(\mathbf{Z})\\
y=\bE_{\mathbf{Z},\dist 2}\fh 2(\mathbf{Z})
}
}=O(1).
\]
By applying similar techniques as in the proofs of Lemmas \ref{lem:exp_bias}
and \ref{lem:exp_bias2}, it can then be shown with the application
of Lemma \ref{lem:var_term} and the fact that with high probability
\[
\left(\frac{1}{\Pr\left(S_{k_{i}}(\mathbf{Z})\right)}\frac{k_{i}-1}{N_{i}}\right)^{2}=O\left(\frac{1}{k_{i}}\right),
\]
the expected value of (\ref{eq:gvariance}) reduces to 
\begin{equation}
\bE\left[g\left(\bE_{\mathbf{Z},\dist 1(\mathbf{Z})}\fh 1(\mathbf{Z}),\bE_{\mathbf{Z},\dist 2(\mathbf{Z})}\fh 2(\mathbf{Z})\right)\right]+O\left(\frac{1}{k_{1}}+\frac{1}{k_{2}}\right).\label{eq:var_term1}
\end{equation}

If $g(x,y)$ has mixed derivatives of the form of $x^{\alpha}y^{\beta}$
for $\alpha,\beta\in\mathbb{R}$, we can apply the generalized binomial
theorem prior to taking the expectation to show that 
\begin{eqnarray}
\lefteqn{\bE\left[g\left(\fh 1(\mathbf{Z}),\fh 2(\mathbf{Z})\right)-g\left(\bE_{\mathbf{Z},\dist 1(\mathbf{Z})}\fh 1(\mathbf{Z}),\bE_{\mathbf{Z},\dist 2(\mathbf{Z})}\fh 2(\mathbf{Z})\right)\right]}\nonumber \\
 & = & \sum_{j=1}^{\nu/2}\sum_{m=0}^{r}\sum_{i=1}^{\nu/2}\sum_{n=0}^{r}\frac{c_{19,j,i,m,n}}{k_{1}^{j}k_{2}^{i}}\left(\frac{k_{1}}{N_{1}}\right)^{\frac{m}{d}}\left(\frac{k_{2}}{N_{2}}\right)^{\frac{n}{d}}+O\left(\frac{1}{k_{1}^{\nu/2}}+\frac{1}{k_{2}^{\nu/2}}+\left(\frac{k_{1}}{N_{1}}\right)^{\frac{s}{d}}+\left(\frac{k_{2}}{N_{2}}\right)^{\frac{s}{d}}\right)\nonumber \\
 &  & +\sum_{j=1}^{\nu/2}\sum_{m=0}^{r}\left(\frac{c_{19,1,j,m}}{k_{1}^{j}}\left(\frac{k_{1}}{N_{1}}\right)^{\frac{m}{d}}+\frac{c_{19,2,j,m}}{k_{2}^{j}}\left(\frac{k_{2}}{N_{2}}\right)^{\frac{m}{d}}\right).\label{eq:var_term2}
\end{eqnarray}
Combining (\ref{eq:bias_term}) with either (\ref{eq:var_term1})
or (\ref{eq:var_term2}) completes the proof of Theorem \ref{thm:biasKNN}.

\section{Proof of Theorem~\ref{thm:varianceKNN} (Variance)}

\label{sec:varProofKNN}To bound the variance of the plug-in estimator
$\ghat$, we will again use the Efron-Stein inequality~\cite{efron1981jackknife}:
\begin{lem}
[Efron-Stein Inequality] \label{lem:efron} Let $\mathbf{X}_{1},\dots,\mathbf{X}_{n},\mathbf{X}_{1}^{'},\dots,\mathbf{X}_{n}^{'}$
be independent random variables on the space $\mathcal{S}$. Then
if $f:\mathcal{S}\times\dots\times\mathcal{S}\rightarrow\mathbb{R}$,
we have that 
\[
\var\left[f(\mathbf{X}_{1},\dots,\mathbf{X}_{n})\right]\leq\frac{1}{2}\sum_{i=1}^{n}\bE\left[\left(f(\mathbf{X}_{1},\dots,\mathbf{X}_{n})-f(\mathbf{X}_{1},\dots,\mathbf{X}_{i}^{'},\dots,\mathbf{X}_{n})\right)^{2}\right].
\]
\end{lem}
Suppose we have samples $\left\{ \mathbf{X}_{1},\dots,\mathbf{X}_{N_{2}},\mathbf{Y}_{1},\dots,\mathbf{Y}_{N_{1}}\right\} $
and $\left\{ \mathbf{X}_{1}^{'},\dots,\mathbf{X}_{N_{2}},\mathbf{Y}_{1},\dots,\mathbf{Y}_{N_{1}}\right\} $
and denote the respective estimators as $\ghat$ and $\ghat^{'}$.
We have that 
\begin{align}
\left|\ghat-\ghat^{'}\right| & \leq & \frac{1}{N_{2}}\left|g\left(\fh 1(\mathbf{X}_{1}),\fh 2(\mathbf{X}_{1})\right)-g\left(\fh 1(\mathbf{X}_{1}^{'}),\fh 2(\mathbf{X}_{1}^{'})\right)\right|\nonumber \\
 &  & +\frac{1}{N_{2}}\sum_{j=2}^{N_{2}}\left|g\left(\fh 1(\mathbf{X}_{j}),\fh 2(\mathbf{X}_{j})\right)-g\left(\fh 1(\mathbf{X}_{j}),\fh 2^{'}(\mathbf{X}_{j})\right)\right|.\label{eq:triangle}
\end{align}

Define $\mathbf{P}_{k_{i}}(\mathbf{X}_{j})=\Pr\left(S_{k_{i}}(\mathbf{X}_{j})\right)$.
This is a random variable denoting the probability that a point drawn
from $f_{i}$ falls into the $k_{i}$th nearest neighbor ball of $\mathbf{X}_{j}$.
As mentioned in Appendix~\ref{sec:biasProofKNN}, the distribution
of $\mathbf{P}_{k_{i}}(\mathbf{X}_{j})$ is independent of $\mathbf{X}_{j}$
and $f_{i}$ and is a beta random variable~\cite{fukunaga1973optimization}
with density 
\[
f_{k_{i}}\left(p_{k_{i}}\right)=\frac{M_{i}!}{\left(k_{i}-1\right)!(M_{i}-k_{i})!}p_{k_{i}}^{k_{i}-1}\left(1-p_{k_{i}}\right)^{M_{i}-k_{i}}.
\]
Define 
\[
\fb i(\mathbf{X}_{j})=f_{i}(\mathbf{X}_{j})\frac{k_{i}-1}{M_{i}\mathbf{P}_{k_{i}}(\mathbf{X}_{j})}.
\]
We then have that with high probability \cite{sricharan2012thesis},
\begin{equation}
\fh i(\mathbf{X}_{j})=\fb i(\mathbf{X}_{j})+O\left(\left(\frac{k_{i}}{M_{i}}\right)^{\frac{2}{d}}\right).\label{eq:beta}
\end{equation}

The following lemma can be used to control the first term in (\ref{eq:triangle}):
\begin{lem}
\label{lem:1stterm_final} 
\[
\bE\left[\left|g\left(\fh 1(\mathbf{X}_{1}),\fh 2(\mathbf{X}_{1})\right)-g\left(\fh 1(\mathbf{X}_{1}^{'}),\fh 2(\mathbf{X}_{1}^{'})\right)\right|^{2}\right]=O(1).
\]
\end{lem}
\begin{IEEEproof}
Since $g$ is Lipschitz continuous with constant $C_{g}$, we have
\begin{eqnarray}
\left|g\left(\fh 1(\mathbf{X}_{1}),\fh 2(\mathbf{X}_{1})\right)-g\left(\fh 1(\mathbf{X}_{1}^{'}),\fh 2(\mathbf{X}_{1}^{'})\right)\right| & \leq & C_{g}\left|\fh 1(\mathbf{X}_{1})-\fh 1(\mathbf{X}_{1}^{'})\right|\nonumber \\
 &  & +C_{g}\left|\fh 2(\mathbf{X}_{1})-\fh 2(\mathbf{X}_{1}^{'})\right|.\label{eq:lipschitz}
\end{eqnarray}
From the triangle inequality, Jensen's inequality, and (\ref{eq:beta}),
we get 
\begin{eqnarray}
\bE\left[\left|\ft i(\mathbf{X}_{1})-\ft i(\mathbf{X}_{1}^{'})\right|^{2}\right] & \leq & 2\bE\left[\left(\fh i(\mathbf{X}_{1})\right)^{2}\right]\nonumber \\
 & \leq & 4\bE\left[\left(\fb i(\mathbf{X}_{1})\right)^{2}\right]+O\left(\left(\frac{k_{i}}{M_{i}}\right)^{\frac{4}{d}}\right)\nonumber \\
 & = & 4\bE\left[f_{i}^{2}(\mathbf{X}_{1})\right]\frac{\left(k_{i}-1\right)^{2}}{M_{i}^{2}}\cdot\frac{M_{i}(M_{i}-1)}{(k_{i}-1)(k_{i}-2)}+O\left(\left(\frac{k_{i}}{M_{i}}\right)^{\frac{4}{d}}\right).\label{eq:densityBound}
\end{eqnarray}
Combining (\ref{eq:densityBound}) with (\ref{eq:lipschitz}) after
applying Jensen's inequality gives the result. 
\end{IEEEproof}
To control the second term in (\ref{eq:triangle}), consider the following
events: 
\begin{itemize}
\item $A_{1}(\mathbf{X}_{i})$: $\mathbf{X}_{1}$ is strictly within the
$k_{2}$-nn ball around $\mathbf{X}_{i}$ wrt the sample $\left\{ \mathbf{X}_{1},\dots,\mathbf{X}_{N_{2}}\right\} \backslash\{\mathbf{X}_{i}\}$. 
\item $A_{2}(\mathbf{X}_{i})$: $\mathbf{X}_{1}$ is the $k_{2}$th nearest
neighbor of $\mathbf{X}_{i}$ wrt the sample $\left\{ \mathbf{X}_{1},\dots,\mathbf{X}_{N_{2}}\right\} \backslash\{\mathbf{X}_{i}\}$. 
\item $A_{3}(\mathbf{X}_{i})$: $\mathbf{X}_{1}$ is strictly outside of
the $k_{2}$-nn ball around $\mathbf{X}_{i}$ wrt the sample $\left\{ \mathbf{X}_{1},\dots,\mathbf{X}_{N_{2}}\right\} \backslash\{\mathbf{X}_{i}\}$. 
\item $B_{1}(\mathbf{X}_{i})$: $\mathbf{X}_{1}^{'}$ is strictly within
the $k_{2}$-nn ball around $\mathbf{X}_{i}$ wrt the sample \linebreak{}
$\left\{ \mathbf{X}_{1}^{'},\mathbf{X}_{2},\dots,\mathbf{X}_{N_{2}}\right\} \backslash\{\mathbf{X}_{i}\}$. 
\item $B_{2}(\mathbf{X}_{i})$: $\mathbf{X}_{1}^{'}$ is the $k_{2}$th
nearest neighbor of $\mathbf{X}_{i}$ wrt the sample $\left\{ \mathbf{X}_{1}^{'},\mathbf{X}_{2},\dots,\mathbf{X}_{N_{2}}\right\} \backslash\{\mathbf{X}_{i}\}$. 
\item $B_{3}(\mathbf{X}_{i}):$ $\mathbf{X}_{1}^{'}$ is strictly outside
the $k_{2}$-nn ball around $\mathbf{X}_{i}$ wrt the sample \linebreak{}
$\left\{ \mathbf{X}_{1}^{'},\mathbf{X}_{2},\dots,\mathbf{X}_{N_{2}}\right\} \backslash\{\mathbf{X}_{i}\}$. 
\item $BE(\mathbf{X}_{i})=\left(A_{1}(\mathbf{X}_{i})\cap B_{3}(\mathbf{X}_{i})\right)\cup\left(A_{3}(\mathbf{X}_{i})\cap B_{1}(\mathbf{X}_{i})\right)$. 
\item $BE_{1}(\mathbf{X}_{i},\mathbf{X}_{j})=BE(\mathbf{X}_{i})\cap\left[BE(\mathbf{X}_{j})\cup A_{2}(\mathbf{X}_{j})\cup B_{2}(\mathbf{X}_{j})\right]$. 
\item \textbf{$BE_{2}(\mathbf{X}_{i},\mathbf{X}_{j})=A_{2}(\mathbf{X}_{i})\cap\left[A_{2}(\mathbf{X}_{j})\cup B_{2}(\mathbf{X}_{j})\right]$.} 
\item $BE_{3}(\mathbf{X}_{i},\mathbf{X}_{j})=B_{2}(\mathbf{X}_{i})\cap B_{2}(\mathbf{X}_{j})$. 
\end{itemize}
Note that if neither $BE_{1}(\mathbf{X}_{i},\mathbf{X}_{j})$, $BE_{2}(\mathbf{X}_{i},\mathbf{X}_{j})$,
nor $BE_{3}(\mathbf{X}_{i},\mathbf{X}_{j})$ hold, then 
\begin{eqnarray}
\left|g\left(\fh 1(\mathbf{X}_{i}),\fh 2(\mathbf{X}_{i})\right)-g\left(\fh 1(\mathbf{X}_{i}),\fh 2^{'}(\mathbf{X}_{i})\right)\right|\left|g\left(\fh 1(\mathbf{X}_{i}),\fh 2(\mathbf{X}_{i})\right)-g\left(\fh 1(\mathbf{X}_{i}),\fh 2^{'}(\mathbf{X}_{i})\right)\right| & = & 0,\label{eq:knn_same}
\end{eqnarray}
since either $\fh 2^{'}(\mathbf{X}_{i})=\fh 2(\mathbf{X}_{i})$ or
$\fh 2^{'}(\mathbf{X}_{j})=\fh 2(\mathbf{X}_{j})$. The same result
holds if $\mathbf{X}_{i}$ or $\mathbf{X}_{j}$ are switched. Thus
we only need to focus on the cases where these events are true. Note
that since the samples are iid, the probability that $A_{2}(\mathbf{X}_{i})$
occurs is $1/N_{2}$. Similarly, the probability of $B_{2}(\mathbf{X}_{i})$
is $1/N_{2}$.
\begin{claim}
\label{cla:probability}The following hold: 
\begin{enumerate}
\item $\Pr\left(BE_{1}(\mathbf{X}_{i},\mathbf{X}_{j})\right)=O\left(\left(\frac{k_{2}}{M_{2}}\right)^{2}\right)$ 
\item $\Pr\left(BE_{2}(\mathbf{X}_{i},\mathbf{X}_{j})\right)=O\left(\frac{1}{N_{2}^{2}}\right)$ 
\item $\Pr\left(BE_{3}(\mathbf{X}_{i},\mathbf{X}_{j})\right)=O\left(\frac{1}{N_{2}^{2}}\right)$ 
\end{enumerate}
\end{claim}
\begin{IEEEproof}
For the first expression, consider first the case $BE(\mathbf{X}_{i})\cap BE(\mathbf{X}_{j})$.
If $\mathbf{X}_{i}$ and $\mathbf{X}_{j}$ are far apart with disjoint
$k_{2}$-nn balls, we can treat the probability of $BE(\mathbf{X}_{i})$
and $BE(\mathbf{X}_{j})$ separately within each ball which is $O\left(\frac{k_{2}}{M_{2}}\right)$
in each case. This gives a combined probability of $O\left(\left(\frac{k_{2}}{M_{2}}\right)^{2}\right)$
when the balls are disjoint. On the other hand, the probability that
the $k_{2}$-nn balls intersect is $O\left(\frac{k_{2}}{M_{2}}\right)$.
In this case, the probability of the event $BE(\mathbf{X}_{i})\cap BE(\mathbf{X}_{j})$
is $O\left(\frac{k_{2}}{M_{2}}\right)$. Combining these facts proves
the claim for $BE(\mathbf{X}_{i})\cap BE(\mathbf{X}_{j})$.

Now consider $BE(\mathbf{X}_{i})\cap A_{2}(\mathbf{X}_{j})$. In a
similar manner as above, if the two $k_{2}$-nn balls are disjoint,
we treat the probability of the two events separately within each
ball separately giving a combined probability of $O\left(\frac{k_{2}}{M_{2}^{2}}\right)$.
Again, the probability that the $k_{2}$-nn balls intersect is $O\left(\frac{k_{2}}{M_{2}}\right)$
and the resulting probability of $BE(\mathbf{X}_{i})\cap A_{2}(\mathbf{X}_{j})$
is $O\left(\frac{k_{2}}{M_{2}}\right)$ giving a combined probability
of $O\left(\left(\frac{k_{2}}{M_{2}}\right)^{2}\right).$ Similarly,
$\Pr\left(BE(\mathbf{X}_{i})\cap B_{2}(\mathbf{X}_{j})\right)=O\left(\left(\frac{k_{2}}{M_{2}}\right)^{2}\right)$
which completes the proof for the first expression.

For the second and third expressions, note that since the points $\left\{ \mathbf{X}_{1}^{'},\mathbf{X}_{1},\mathbf{X}_{2},\dots,\mathbf{X}_{N_{2}}\right\} $
are all iid, $A_{2}(\mathbf{X}_{i})$ is independent of $A_{2}(\mathbf{X}_{j})$
and $B_{2}(\mathbf{X}_{j}$) and $B_{2}(\mathbf{X}_{i})$ is independent
of $B_{2}(\mathbf{X}_{j})$. Thus the probability of each of the intersecting
events is $1/N_{2}^{2}$ which completes the proof.
\end{IEEEproof}
From the Lipschitz condition, 
\begin{align}
\left|g\left(\fh 1(\mathbf{X}_{j}),\fh 2(\mathbf{X}_{j})\right)-g\left(\fh 1(\mathbf{X}_{j}),\fh 2^{'}(\mathbf{X}_{j})\right)\right|^{2} & \leq & C_{g}^{2}\left|\fh 2(\mathbf{X}_{j})-\fh 2^{'}(\mathbf{X}_{j})\right|^{2}\label{eq:lipschitz2}
\end{align}
Now suppose that $A_{1}(\mathbf{X}_{j})\cap B_{3}(\mathbf{X}_{j})$
occurs. In this case, $\fh 2^{'}(\mathbf{X}_{j})=\frac{k_{2}-1}{k_{2}}\fbp 2(\mathbf{X}_{j})$.
To obtain a bound for $\bE\left[\left|\fh 2(\mathbf{X}_{j})-\fh 2^{'}(\mathbf{X}_{j})\right|^{2}\right]$,
we need the joint distribution of $\fb 2(\mathbf{X}_{j})$ and $\fbp 2(\mathbf{X}_{j})$
as 
\begin{equation}
\left|\fh 2(\mathbf{X}_{j})-\fh 2^{'}(\mathbf{X}_{j})\right|^{2}\leq2\left|\fb 2(\mathbf{X}_{j})-\frac{k_{2}-1}{k_{2}}\fbp 2(\mathbf{X}_{j})\right|^{2}+O\left(\left(\frac{k_{2}}{M_{2}}\right)^{\frac{4}{d}}\right).\label{eq:fhat2fbar}
\end{equation}

\begin{lem}
\label{lem:joint_density}The density function of the joint distribution
of $\mathbf{P}_{k_{2}}$ and $\mathbf{P}_{k_{2}+1}$ is\textup{\emph{
\begin{equation}
f_{P_{k_{2}},P_{k_{2}+1}}\left(p,q\right)=1_{\{p\leq q\}}\frac{M_{2}!}{\left(k_{2}-1\right)!\left(M_{2}-k_{2}-1\right)!}p^{k_{2}-1}\left(1-q\right)^{M_{2}-k_{2}-1}.\label{eq:joint}
\end{equation}
}}
\end{lem}
\begin{IEEEproof}
For $\mathbf{P}_{k_{2}}$, let $\mathbf{r}_{k_{2}}$ be the corresponding
$k$-nn radius. Let $\delta_{p}$, $\delta_{q}>0$. We are interested
in the event $\left\{ p\leq\mathbf{P}_{k_{2}}\leq p+\delta_{p},q\leq\mathbf{P}_{k_{2}+1}\leq q+\delta_{q}\right\} $.
Consider the following events: 
\begin{itemize}
\item $C_{1}$: There are $k_{2}-1$ points within the radius $\mathbf{r}_{k_{2}}$. 
\item $C_{2}$: The $k_{2}$th point is in the interval $[\mathbf{r}_{k_{2}},\mathbf{r}_{k_{2}}+\epsilon(\delta_{p})]$. 
\item $C_{3}$: The $k_{2}+1$th point is in the interval $[\mathbf{r}_{k_{2}+1},\mathbf{r}_{k_{2}+1}+\epsilon(\delta_{q})]$. 
\item $C_{4}$: The remaining $M_{2}-k_{2}-1$ points are outside the radius
$\mathbf{r}_{k_{2}+1}+\epsilon(\delta_{q})$. 
\item $C_{5}$: $\mathbf{r}_{k_{2}}\leq\mathbf{r}_{k_{2}+1}$ 
\end{itemize}
We have that 
\[
\Pr\left(p\leq\mathbf{P}_{k_{2}}\leq p+\delta_{p},q\leq\mathbf{P}_{k_{2}+1}\leq q+\delta_{q}\right)=\Pr\left(\bigcap_{i=1}^{5}C_{i}\right).
\]
Of the $M_{2}!$ different ways to permute the $M_{2}$ points, there
are $(k_{2}-1)!$ permutations for the points inside the $k_{2}$-nn
ball and $(M_{2}-k_{2}-1)!$ permutations for the points outside the
$(k_{2}+1)$-nn ball. So the number of different point configurations
with $k_{2}-1$ points inside $\mathbf{r}_{k_{2}}$ and $M_{2}-k_{2}-1$
points outside $\mathbf{r}_{k_{2}+1}$ is $\frac{M_{2}!}{(k_{2}-1)!(M_{2}-k_{2}-1)!}$.
This gives 
\begin{align}
\lefteqn{\Pr\left(p\leq\mathbf{P}_{k_{2}}\leq p+\delta_{p},q\leq\mathbf{P}_{k_{2}+1}\leq q+\delta_{q}\right)}\nonumber \\
 & = & 1_{\{p\leq q\}}\frac{M_{2}!}{(k_{2}-1)!(M_{2}-k_{2}-1)!}p^{k_{2}-1}(1-q)^{M_{2}-k_{2}-1}\delta_{p}\delta_{q}.\label{eq:joint_pre}
\end{align}
The $p^{k_{2}-1}$ term is the probability that $k_{2}-1$ points
fall within a ball of radius $p$ (the coverage probability). The
$(1-q)^{M_{2}-k_{2}-1}$ term is the probability that $M_{2}-k_{2}-1$
points fall outside a ball of radius with coverage probability $q$.
The $\delta_{q}$ and $\delta_{p}$ terms correspond to the events
that one point falls exactly at radius $p$ and another point falls
exactly at radius $q$. The LHS of (\ref{eq:joint_pre}) is equal
to the probability of these events. The combinatorial term then accurately
accounts for the different possible combinations. From (\ref{eq:joint_pre}),
we get the density in (\ref{eq:joint}).
\end{IEEEproof}
From Lemma~\ref{lem:joint_density}, 
\begin{eqnarray*}
\bE\left[\fb 2(\mathbf{X}_{j})\fbp 2(\mathbf{X}_{j})\right] & = & \bE\left[f_{2}^{2}(\mathbf{X}_{j})\frac{k_{2}(k_{2}-1)}{M_{2}^{2}\mathbf{P}_{k_{2}}(\mathbf{X}_{j})\mathbf{P}_{k_{2}+1}(\mathbf{X}_{j})}\right]\\
 & = & \bE\left[f_{2}^{2}(\mathbf{X}_{j})\right]\frac{k_{2}(M_{2}-1)}{(k_{2}-1)M_{2}}.
\end{eqnarray*}
Then since $\bE\left[\mathbf{P}_{k_{2}}^{-2}\right]=\frac{M_{2}(M_{2}-1)}{(k_{2}-1)(k_{2}-2)}$,
we obtain 
\begin{eqnarray}
\bE\left[\left|\fb 2(\mathbf{X}_{j})-\frac{k_{2}-1}{k_{2}}\fbp 2(\mathbf{X}_{j})\right|^{2}\right] & = & \bE\left[f_{2}^{2}(\mathbf{X}_{j})\right]\frac{M_{2}-1}{M_{2}}\cdot\frac{2}{k_{2}\left(k_{2}-2\right)}\nonumber \\
 & = & O\left(\frac{1}{k_{2}^{2}}\right).\label{eq:fbar_diff}
\end{eqnarray}
A similar result follows if $A_{3}(\mathbf{X}_{i})\cap B_{1}(\mathbf{X}_{i})$
holds instead. Then (\ref{eq:knn_same}) gives 
\begin{align}
\lefteqn{\bE\left[\left(\sum_{j=2}^{N_{2}}\left|g\left(\fh 1(\mathbf{X}_{j}),\fh 2(\mathbf{X}_{j})\right)-g\left(\fh 1(\mathbf{X}_{j}),\fh 2^{'}(\mathbf{X}_{j})\right)\right|\right)^{2}\right]}\nonumber \\
 & = & \sum_{i=2}^{N_{2}}\sum_{j=2}^{N_{2}}\bE\left[\left|g\left(\fh 1(\mathbf{X}_{i}),\fh 2(\mathbf{X}_{i})\right)-g\left(\fh 1(\mathbf{X}_{i}),\fh 2^{'}(\mathbf{X}_{i})\right)\right|\right.\nonumber \\
 &  & \times\left.\left|g\left(\fh 1(\mathbf{X}_{j}),\fh 2(\mathbf{X}_{j})\right)-g\left(\fh 1(\mathbf{X}_{j}),\fh 2^{'}(\mathbf{X}_{j})\right)\right|\right]\nonumber \\
 & \leq & \sum_{i=2}^{N_{2}}\sum_{j=2}^{N_{2}}2\bE\left[\left|g\left(\fh 1(\mathbf{X}_{i}),\fh 2(\mathbf{X}_{i})\right)-g\left(\fh 1(\mathbf{X}_{i}),\fh 2^{'}(\mathbf{X}_{i})\right)\right|\right.\nonumber \\
 &  & \times\left.\left.\left|g\left(\fh 1(\mathbf{X}_{j}),\fh 2(\mathbf{X}_{j})\right)-g\left(\fh 1(\mathbf{X}_{j}),\fh 2^{'}(\mathbf{X}_{j})\right)\right|\right|\bigcup_{\ell=1}^{3}BE_{\ell}(\mathbf{X}_{i},\mathbf{X}_{j})\right]\nonumber \\
 &  & \times\Pr\left(\bigcup_{\ell=1}^{3}BE_{\ell}(\mathbf{X}_{i},\mathbf{X}_{j})\right)\label{eq:square_sum-1}
\end{align}
Combining the results from (\ref{eq:square_sum-1}), (\ref{eq:lipschitz2}),
(\ref{eq:fhat2fbar}), (\ref{eq:fbar_diff}), and Claim~\ref{cla:probability}
with the Cauchy-Schwarz inequality gives 
\begin{eqnarray}
\lefteqn{\text{LHS (\ref{eq:square_sum-1})}}\nonumber \\
 & \leq & 2M_{2}^{2}\bE\left[\left.\left|g\left(\fh 1(\mathbf{X}_{i}),\fh 2(\mathbf{X}_{i})\right)-g\left(\fh 1(\mathbf{X}_{i}),\fh 2^{'}(\mathbf{X}_{i})\right)\right|^{2}\right|BE_{1}(\mathbf{X}_{i},\mathbf{X}_{j})\right]\nonumber \\
 &  & \times\Pr\left(BE_{1}(\mathbf{X}_{i},\mathbf{X}_{j})\right)+O\left(\frac{M_{2}^{2}}{N_{2}^{2}}\right)\nonumber \\
 & \leq & 2M_{2}^{2}C_{g}^{2}\bE\left[\left.\left|\fh 2(\mathbf{X}_{j})-\fh 2^{'}(\mathbf{X}_{j})\right|^{2}\right|BE_{1}(\mathbf{X}_{i},\mathbf{X}_{j})\right]\Pr\left(BE_{1}(\mathbf{X}_{i},\mathbf{X}_{j})\right)+O(1)\nonumber \\
 & \leq & 4M_{2}^{2}C_{g}^{2}\bE\left[\left.\left|\fb 2(\mathbf{X}_{j})-\frac{k_{2}-1}{k_{2}}\fbp 2(\mathbf{X}_{j})\right|^{2}\right|BE_{1}(\mathbf{X}_{i},\mathbf{X}_{j})\right]\Pr\left(BE_{1}(\mathbf{X}_{i},\mathbf{X}_{j})\right)\nonumber \\
 &  & +O\left(\left(\frac{k_{2}}{M_{2}}\right)^{\frac{4}{d}}+1\right)\nonumber \\
 & = & O\left(M_{2}^{2}\cdot\frac{1}{k_{2}^{2}}\cdot\left(\frac{k_{2}}{M_{2}}\right)^{2}\right)+O\left(\left(\frac{k_{2}}{M_{2}}\right)^{\frac{4}{d}}+1\right)\nonumber \\
 & = & O(1).\label{eq:2ndtermFinal}
\end{eqnarray}

Applying Jensen's inequality to (\ref{eq:triangle}) and applying
(\ref{eq:2ndtermFinal}) and Lemma~\ref{lem:1stterm_final} gives
\begin{eqnarray*}
\lefteqn{\bE\left[\left|\ghat-\ghat^{'}\right|^{2}\right]}\\
 & \leq & \frac{2}{N_{2}^{2}}\bE\left[\left|g\left(\fh 1(\mathbf{X}_{1}),\fh 2(\mathbf{X}_{1})\right)-g\left(\fh 1(\mathbf{X}_{1}^{'}),\fh 2(\mathbf{X}_{1}^{'})\right)\right|^{2}\right]\\
 &  & +\frac{2}{N_{2}^{2}}\bE\left[\left(\sum_{j=2}^{N_{2}}\left|g\left(\fh 1(\mathbf{X}_{j}),\fh 2(\mathbf{X}_{j})\right)-g\left(\fh 1(\mathbf{X}_{j}),\fh 2^{'}(\mathbf{X}_{j})\right)\right|\right)^{2}\right]\\
 & = & O\left(\frac{1}{N_{2}^{2}}\right).
\end{eqnarray*}

Now suppose we have samples $\left\{ \mathbf{X}_{1},\dots,\mathbf{X}_{N_{2}},\mathbf{Y}_{1},\dots,\mathbf{Y}_{N_{1}}\right\} $
and $\left\{ \mathbf{X}_{1},\dots,\mathbf{X}_{N_{2}},\mathbf{Y}_{1}^{'},\dots,\mathbf{Y}_{N_{1}}\right\} $
and denote the respective estimators as $\ghat$ and $\ghat^{'}$.
Then 
\begin{eqnarray*}
\left|g\left(\fh 1(\mathbf{X}_{j}),\fh 2(\mathbf{X}_{j})\right)-g\left(\fh 1^{'}(\mathbf{X}_{j}),\fh 2(\mathbf{X}_{j})\right)\right| & \leq & C_{g}\left|\fh 1(\mathbf{X}_{j})-\fh 1^{'}(\mathbf{X}_{j})\right|
\end{eqnarray*}
Thus by similar arguments as was used to obtain (\ref{eq:2ndtermFinal}),
\begin{eqnarray*}
\lefteqn{\bE\left[\left|\ghat-\ghat^{'}\right|^{2}\right]}\\
 & \leq & \frac{1}{N_{2}^{2}}\bE\left[\left(\sum_{j=1}^{N_{2}}\left|g\left(\fh 1(\mathbf{X}_{j}),\fh 2(\mathbf{X}_{j})\right)-g\left(\fh 1^{'}(\mathbf{X}_{j}),\fh 2(\mathbf{X}_{j})\right)\right|\right)^{2}\right]\\
 & = & O\left(\frac{1}{N_{2}^{2}}\right).
\end{eqnarray*}
Applying the Efron-Stein inequality gives 
\[
\var\left[\ghat\right]=O\left(\frac{1}{N_{2}}+\frac{N_{1}}{N_{2}^{2}}\right).
\]

\section{Proof of Theorem~\ref{thm:cltKNN} (CLT)}

\label{sec:cltProofKNN}We use Lemma~\ref{lem:clt_covariance} which
is adapted from~\cite{sricharan2012estimation}:
\begin{lem}
\label{lem:clt_covariance}Let the random variables $\{\mathbf{Y}_{M,i}\}_{i=1}^{N}$
belong to a zero mean, unit variance, interchangeable process for
all values of $M$. Assume that $Cov(\mathbf{Y}_{M,1},\mathbf{Y}_{M,2})$
and $Cov(\mathbf{Y}_{M,1}^{2},\mathbf{Y}_{M,2}^{2})$ are $o(1)$
as $M\rightarrow\infty$. Then the random variable 
\begin{equation}
\mathbf{S}_{N,M}=\frac{\sum_{i=1}^{N}\mathbf{Y}_{M,i}}{\sqrt{\var\left[\sum_{i=1}^{N}\mathbf{Y}_{M,i}\right]}}\label{eq:sum-1}
\end{equation}
converges in distribution to a standard normal random variable.
\end{lem}
The proof of this lemma is identical to that in \cite{sricharan2012estimation}
(See ``Proof of Theorem 3.3 and Theorem 5.3'' in \cite{sricharan2012estimation}).
The relaxed assumptions in Lemma~\ref{lem:clt_covariance} enable
us to prove the central limit theorem under more relaxed conditions
on the densities. Assume for simplicity that $N_{1}=M_{2}=M$ and
$k_{1}(l)=k_{2}(l)=k(l)$. Define 
\begin{equation}
\mathbf{Y}_{M,i}=\frac{\sum_{l\in\bar{l}}w(l)g\left(\fhl 1l\mathbf{X}_{i}),\fhl 2l\mathbf{X}_{i}\right)-\bE\left[\sum_{l\in\bar{l}}w(l)g\left(\fhl 1l\mathbf{X}_{i}),\fhl 2l\mathbf{X}_{i}\right)\right]}{\sqrt{\var\left[\sum_{l\in\bar{l}}w(l)g\left(\fhl 1l\mathbf{X}_{i}),\fhl 2l\mathbf{X}_{i}\right)\right]}}.\label{eq:Ydef}
\end{equation}
This gives 
\[
\mathbf{S}_{N,M}=\frac{\hat{\mathbf{G}}_{w}-\bE\left[\hat{\mathbf{G}}_{w}\right]}{\sqrt{\var\left[\hat{\mathbf{G}}_{w}\right]}}.
\]

To bound the covariance between $\mathbf{Y}_{M,1}$ and $\mathbf{Y}_{M,2}$
and between $\mathbf{Y}_{M,1}^{2}$ and $\mathbf{Y}_{M,2}^{2}$, it
is necessary to show that the denominator of $\mathbf{Y}_{M,i}$ converges
to a nonzero constant or to zero sufficiently slowly. The numerator
and denominator of $\mathbf{Y}_{M,i}$ are, respectively, 
\[
\sum_{l\in\bar{l}}w(l)g\left(\fhl 1l(\mathbf{X}_{i}),\fhl 2l(\mathbf{X}_{i})\right)-\bE\left[\sum_{l\in\bar{l}}w(l)g\left(\fhl 1l(\mathbf{X}_{i}),\fhl 2l(\mathbf{X}_{i})\right)\right]
\]
\[
=\sum_{l\in\bar{l}}w(l)\left(g\left(\fhl 1l(\mathbf{X}_{i}),\fhl 2l(\mathbf{X}_{i})\right)-\bE\left[g\left(\fhl 1l(\mathbf{X}_{i}),\fhl 2l(\mathbf{X}_{i})\right)\right]\right),
\]
\[
\sqrt{\var\left[\sum_{l\in\bar{l}}w(l)g\left(\fhl 1l(\mathbf{X}_{i}),\fhl 2l(\mathbf{X}_{i})\right)\right]}
\]
\begin{equation}
=\sqrt{\sum_{l\in\bar{l}}\sum_{l'\in\bar{l}}w(l)w(l')Cov\left(g\left(\fhl 1l(\mathbf{X}_{i}),\fhl 2l(\mathbf{X}_{i})\right),g\left(\fhl 1{l'}(\mathbf{X}_{i}),\fhl 2{l'}(\mathbf{X}_{i})\right)\right).}\label{eq:Yden}
\end{equation}
Thus we require bounds on $Cov\left(g\left(\fhl 1l(\mathbf{X}_{i}),\fhl 2l(\mathbf{X}_{i})\right),g\left(\fhl 1{l'}(\mathbf{X}_{j}),\fhl 2{l'}(\mathbf{X}_{j})\right)\right)$
to bound the covariance between $\mathbf{Y}_{M,1}$ and $\mathbf{Y}_{M,2}$.

Define $\mathcal{M}(\mathbf{Z}):=\mathbf{Z}-\mathbb{E}\mathbf{Z}$
and $\ehatl il(\mathbf{Z}):=\fhl il(\mathbf{Z})-\mathbb{E}_{\mathbf{Z}}\fhl il(\mathbf{Z})$.
A Taylor series expansion of $g\left(\fhl 1l(\mathbf{X}_{n}),\fhl 2l(\mathbf{X}_{n})\right)$
around $\mathbb{E}_{\mathbf{X}_{n}}\fhl 1l(\mathbf{X}_{n})$ and $\mathbb{E}_{\mathbf{X}_{n}}\fhl 2l(\mathbf{X}_{n})$
gives 
\begin{eqnarray*}
g\left(\fhl 1l(\mathbf{X}_{n}),\fhl 2l(\mathbf{X}_{n})\right) & = & \sum_{i=0}^{1}\sum_{j=0}^{1}\left(\left.\frac{\partial^{i+j}g(x,y)}{\partial x^{i}\partial y^{j}}\right|_{\substack{x=\mathbb{E}_{\mathbf{X}_{n}}\fhl 1l(\mathbf{X}_{n})\\
y=\mathbb{E}_{\mathbf{X}_{n}}\fhl 2l(\mathbf{X}_{n})
}
}\right)\frac{\ehatl 1l^{i}(\mathbf{X}_{n})\ehatl 2l^{j}(\mathbf{X}_{n})}{i!j!}\\
 &  & +o\left(\ehatl 1l(\mathbf{X}_{n})+\ehatl 2l(\mathbf{X}_{n})+\ehatl 1l(\mathbf{X}_{n})\ehatl 2l(\mathbf{X}_{n})\right)
\end{eqnarray*}
Define 
\begin{eqnarray*}
\mathbf{p}_{n}^{(l)} & := & \mathcal{M}\left(g\left(\bE_{\mathbf{X}_{n}}\fhl 1l(\mathbf{X}_{n}),\bE_{\mathbf{X}_{n}}\fhl 2l(\mathbf{X}_{n})\right)\right),\\
\mathbf{q}_{n}^{(l)} & := & \mathcal{M}\left(\frac{\partial}{\partial x}g\left(\bE_{\mathbf{X}_{n}}\fhl 1l(\mathbf{X}_{n}),\bE_{\mathbf{X}_{n}}\fhl 2l(\mathbf{X}_{n})\right)\ehatl 1l(\mathbf{X}_{n})\right),\\
\mathbf{r}_{n}^{(l)} & := & \mathcal{M}\left(\frac{\partial}{\partial y}g\left(\bE_{\mathbf{X}_{n}}\fhl 1l(\mathbf{X}_{n}),\bE_{\mathbf{X}_{n}}\fhl 2l(\mathbf{X}_{n})\right)\ehatl 2l(\mathbf{X}_{n})\right),\\
\mathbf{s}_{n}^{(l)} & := & \mathcal{M}\left(\frac{\partial^{2}}{\partial x\partial y}g\left(\bE_{\mathbf{X}_{n}}\fhl 1l(\mathbf{X}_{n}),\bE_{\mathbf{X}_{n}}\fhl 2l(\mathbf{X}_{n})\right)\ehatl 1l(\mathbf{X}_{n})\ehatl 2l(\mathbf{X}_{n})\right),\\
\mathbf{t}_{n}^{(l)} & := & \mathcal{M}\left(o\left(\ehatl 1l(\mathbf{X}_{n})+\ehatl 2l(\mathbf{X}_{n})+\ehatl 1l(\mathbf{X}_{n})\ehatl 2l(\mathbf{X}_{n})\right)\right).
\end{eqnarray*}
This gives 
\[
Cov\left(g\left(\fhl 1l(\mathbf{X}_{i}),\fhl 2l(\mathbf{X}_{i})\right),g\left(\fhl 1{l'}(\mathbf{X}_{j}),\fhl 2{l'}(\mathbf{X}_{j})\right)\right)
\]
\begin{equation}
=\bE\left[\left(\mathbf{p}_{i}^{(l)}+\mathbf{q}_{i}^{(l)}+\mathbf{r}_{i}^{(l)}+\mathbf{s}_{i}^{(l)}+\mathbf{t}_{i}^{(l)}\right)\left(\mathbf{p}_{j}^{(l')}+\mathbf{q}_{j}^{(l')}+\mathbf{r}_{j}^{(l')}+\mathbf{s}_{j}^{(l')}+\mathbf{t}_{j}^{(l')}\right)\right].\label{eq:g_cov}
\end{equation}

\begin{lem}
\label{lem:error_cov}Let $l,l'\in\bar{l}$ be fixed and $k(l)\rightarrow\infty$
as $M\rightarrow\infty$ for each $l\in\bar{l}$. Let $\gamma_{1}(x)$
and $\gamma_{2}(x)$ be arbitrary functions with $\sup_{x}|\gamma_{i}(x)|<\infty,$
$i=1,2$. Then if $q+r\geq1$ and $q'+r'\geq1$, 
\[
Cov\left(\gamma_{1}(\mathbf{X}_{i})\ehatl il(\mathbf{X}_{i}),\gamma_{2}(\mathbf{X}_{j})\ehatl i{l'}(\mathbf{X}_{j})\right)=O\left(\frac{1}{\sqrt{k(l)k(l')}}\right),
\]
\[
Cov\left(\gamma_{1}(\mathbf{X}_{i})\ehatl 1l^{q}(\mathbf{X}_{i})\ehatl 2l^{r}(\mathbf{X}_{i}),\gamma_{2}(\mathbf{X}_{j})\ehatl 1{l'}^{q'}(\mathbf{X}_{j})\ehatl 2{l'}^{r'}(\mathbf{X}_{j})\right)=O\left(\frac{1}{\sqrt{k(l)^{q+r}k(l')^{q'+r'}}}\right).
\]
\end{lem}
\begin{IEEEproof}
These results follow from an application of Cauchy-Schwarz and Lemma
\ref{lem:var_term}.
\end{IEEEproof}
\begin{lem}
\label{lem:g_cov}Let $l,l'\in\bar{l}$ be fixed and $k(l)\rightarrow\infty$
as $M\rightarrow\infty$ for each $l\in\bar{l}$. Then 
\[
Cov\left(g\left(\fhl 1l(\mathbf{X}_{i}),\fhl 2l(\mathbf{X}_{i})\right),g\left(\fhl 1{l'}(\mathbf{X}_{j}),\fhl 2{l'}(\mathbf{X}_{j})\right)\right)
\]
\[
=\begin{cases}
\bE\left[\mathbf{p}_{i}^{(l)}\mathbf{p}_{i}^{(l')}\right]+O\left(\frac{1}{\sqrt{k(l)k(l')}}\right), & i=j\\
O\left(\frac{1}{\sqrt{k(l)k(l')}}\right)+o\left(\frac{1}{k(l')}\right), & i\neq j.
\end{cases}
\]
\end{lem}
\begin{IEEEproof}
Consider first $i=j$. Applying Lemma~\ref{lem:error_cov} to (\ref{eq:g_cov})
gives 
\[
Cov\left(g\left(\fhl 1l(\mathbf{X}_{i}),\fhl 2l(\mathbf{X}_{i})\right),g\left(\fhl 1{l'}(\mathbf{X}_{i}),\fhl 2{l'}(\mathbf{X}_{i})\right)\right)=\bE\left[\mathbf{p}_{i}^{(l)}\mathbf{p}_{i}^{(l')}\right]+O\left(\frac{1}{\sqrt{k(l)k(l')}}\right).
\]

When $i\neq j$, $\bE\left[\mathbf{p}_{i}^{(l)}\left(\mathbf{p}_{j}^{(l')}+\mathbf{q}_{j}^{(l')}+\mathbf{r}_{j}^{(l')}+\mathbf{s}_{j}^{(l')}+\mathbf{t}_{j}^{(l')}\right)\right]=0$
since $\mathbf{X}_{i}$ and $\mathbf{X}_{j}$ are independent. A direct
application of Lemma~\ref{lem:error_cov} gives 
\begin{eqnarray*}
\bE\left[\mathbf{q}_{i}^{(l)}\mathbf{q}_{j}^{(l')}\right] & = & O\left(\frac{1}{\sqrt{k(l)k(l')}}\right),\\
\bE\left[\mathbf{q}_{i}^{(l)}\mathbf{r}_{j}^{(l')}\right] & = & O\left(\frac{1}{\sqrt{k(l)k(l')}}\right),\\
\bE\left[\mathbf{q}_{i}^{(l)}\mathbf{s}_{j}^{(l')}\right] & = & O\left(\frac{1}{\sqrt{k(l)k(l')^{2}}}\right),\\
\bE\left[\mathbf{s}_{i}^{(l)}\mathbf{s}_{j}^{(l')}\right] & = & O\left(\frac{1}{k(l)k(l')}\right),\\
\bE\left[\mathbf{s}_{i}^{(l)}\mathbf{r}_{j}^{(l')}\right] & = & O\left(\frac{1}{\sqrt{k(l)^{2}k(l')}}\right),\\
\bE\left[\mathbf{r}_{i}^{(l)}\mathbf{r}_{j}^{(l')}\right] & = & O\left(\frac{1}{\sqrt{k(l)k(l')}}\right).
\end{eqnarray*}
To handle the implicit constants in the $\mathbf{t}_{i}^{(l)}$ terms,
Cauchy-Schwarz can be applied with Lemma~\ref{lem:error_cov} to
get 
\begin{eqnarray*}
\bE\left[\mathbf{q}_{i}^{(l)}\mathbf{t}_{j}^{(l')}\right] & = & o\left(\frac{1}{k(l')}\right),\\
\bE\left[\mathbf{r}_{i}^{(l)}\mathbf{t}_{j}^{(l')}\right] & = & o\left(\frac{1}{k(l')}\right),\\
\bE\left[\mathbf{s}_{i}^{(l)}\mathbf{t}_{j}^{(l')}\right] & = & o\left(\frac{1}{k(l')}\right),\\
\bE\left[\mathbf{t}_{i}^{(l)}\mathbf{t}_{j}^{(l')}\right] & = & o\left(\frac{1}{k(l')}\right).
\end{eqnarray*}
Combining these results with (\ref{eq:g_cov}) completes the proof.
\end{IEEEproof}
Since $\mathbf{p}_{i}(l)=\mathcal{M}\left(g\left(f_{1}(\mathbf{X}_{i}),f_{2}(\mathbf{X}_{i})\right)\right)+o(1)$,
$\bE\left[\mathbf{p}_{i}^{(l)}\mathbf{p}_{i}^{(l')}\right]$ is guaranteed
to be a nonzero constant if 
\begin{equation}
\bE\left[g\left(f_{1}(\mathbf{X}_{i}),f_{2}(\mathbf{X}_{i})\right)^{2}\right]\neq\bE\left[g\left(f_{1}(\mathbf{X}_{i}),f_{2}(\mathbf{X}_{i})\right)\right]^{2}.\label{eq:square_equal}
\end{equation}
In this case, applying Lemma~\ref{lem:g_cov} to (\ref{eq:Ydef})
gives $Cov\left(\mathbf{Y}_{M,1},\mathbf{Y}_{M,2}\right)=o(1)$ as
long as $k(l)\rightarrow\infty$ as $M\rightarrow\infty$ for each
$l\in\bar{l}$. Unfortunately, the condition in (\ref{eq:square_equal})
does not hold for the important case of $f$-divergence functionals
when the densities $f_{1}$ and $f_{2}$ are equal almost everywhere.
However, we still have that the denominator in~(\ref{eq:Ydef}) converges
more slowly to zero than the numerator as long as $k(l),k(l')\rightarrow\infty$
at the same rate for each $l,l'\in\bar{l}$ as the $o\left(\frac{1}{k(l')}\right)$
goes to zero faster than $O\left(\frac{1}{\sqrt{k(l)k(l')}}\right)$.
Thus we still get $Cov\left(\mathbf{Y}_{M,1},\mathbf{Y}_{M,2}\right)=o(1)$
in this case.

For the covariance between $\mathbf{Y}_{M,1}^{2}$ and $\mathbf{Y}_{M,2}^{2}$,
we only need to focus on the numerator terms as the denominator terms
will be similar as before. Thus the numerator of the covariance is
\[
\sum_{l\in\bar{l}}\sum_{l'\in\bar{l}}\sum_{j\in\bar{l}}\sum_{j'\in\bar{l}}Cov\left[\left(\gtay p1l+\gtay q1l+\gtay r1l+\gtay s1l\right)\left(\gtay p1{l'}+\gtay q1{l'}+\gtay r1{l'}+\gtay s1{l'}\right),\right.
\]
\[
\left.\left(\gtay p2j+\gtay q2j+\gtay r2j+\gtay s2j\right)\left(\gtay p2{j'}+\gtay q2{j'}+\gtay r2{j'}+\gtay s2{j'}\right)\right].
\]
If $l=l'$ and $j=j'$, then the previous results apply and we get
$O\left(\frac{1}{M}\right)+o\left(\frac{1}{k(l')}\right)$. For the
general case, the terms with either $\gtay p1l\gtay p1{l'}$ in the
left hand side or $\gtay p2j\gtay p2{j'}$ in the right hand side
are zero due to independence. For the remaining terms, we use the
proof of Lemma 10 in~\cite{moon2014nips}. Under certain conditions,
then for functions $\gamma_{1}(x)$ and $\gamma_{2}(x)$ under the
same assumptions as in Lemma~\ref{lem:error_cov}, 
\[
Cov\left[\gamma_{1}(\mathbf{X}_{1})\ehatl 1l^{s}(\mathbf{X}_{1})\ehatl 2l^{q}(\mathbf{X}_{1})\ehatl 1{l'}^{s'}(\mathbf{X}_{1})\ehatl 2{l'}^{q'}(\mathbf{X}_{1}),\right.
\]
\[
\left.\gamma_{2}(\mathbf{X}_{2})\ehatl 1j^{t}(\mathbf{X}_{2})\ehatl 2j^{r}(\mathbf{X}_{2})\ehatl 1{j'}^{t'}(\mathbf{X}_{2})\ehatl 1{j'}^{r'}(\mathbf{X}_{2})\right]
\]
\begin{equation}
=O\left(\frac{1}{k(l)^{\frac{s+q}{2}}k(l')^{\frac{s'+q'}{2}}k(j)^{\frac{t+r}{2}}k(j')^{\frac{t'+r'}{2}}}\right).\label{eq:big_cov}
\end{equation}
As stated in~\cite{moon2014nips}, the conditions required for this
expression to hold are ``(1) There must be at least one positive
exponent on both sides of the arguments in the covariance. (2) $\{s+s'+t+t'\neq1\}\cap\{q+q'+r+r'\neq1\}$.''
If neither of the conditions holds in condition (2), then the covariance
in (\ref{eq:big_cov}) reduces to the covariance with only one error
term on each side. If only one of the conditions holds, then the covariance
is zero. This means that if $k(l),k(l')\rightarrow\infty$ at the
same rate for each $l,l'\in\bar{l}$, then (\ref{eq:big_cov}) reduces
to $o\left(\frac{1}{k(l)}\right)$. Combining this result with the
previous result on the denominator of $\mathbf{Y}_{M,i}$ gives that
$Cov\left(\mathbf{Y}_{M,1}^{2},\mathbf{Y}_{M,2}^{2}\right)=o(1)$.
Then by Lemma~\ref{lem:clt_covariance}, $\frac{\hat{\mathbf{G}}_{w}-\bE\left[\hat{\mathbf{G}}_{w}\right]}{\sqrt{\var\left[\hat{\mathbf{G}}_{w}\right]}}$
converges in distribution to a standard normal random variable. 
\end{document}